\documentclass[]{aastex61}
\usepackage{amsmath, mathtools}
\usepackage{natbib}
\usepackage{multirow}
\usepackage{comment}
\usepackage{blindtext}
\DeclareMathOperator{\sinc}{sinc}
\accepted{October 23, 2019}
\shorttitle{Extracting  21~cm  EoR signal using  drift scans}
\shortauthors{Patwa \& Sethi}

\def \RRI { \affiliation{Raman Research Institute, C.~V.~Raman Avenue, Sadashivanagar, Bengaluru 560080, India} }

\begin{document}

\title{Detecting  21~cm  EoR signal using drift scans:  correlation of time-ordered  visibilities}

\author{Akash Kumar Patwa}\RRI
\author{Shiv Sethi}\RRI

\correspondingauthor{A.~K.~Patwa}
\email{akpatwa@rri.res.in}


\begin{abstract}
We present a formalism to extract the EoR HI power spectrum for drift scans using radio interferometers. Our main aim is to determine the coherence time scale of time-ordered visibilities. We compute the two-point correlation function of the HI visibilities measured at different times to address this question. We determine, for a given baseline, the decorrelation of the amplitude and the phase of this complex function. Our analysis uses primary beams of four ongoing and future interferometers---PAPER, MWA, HERA, and SKA1-Low. We identify physical processes responsible for the decorrelation of the HI signal and isolate their impact by making suitable analytic approximations. The decorrelation time scale  of the amplitude of the correlation function lies in the range of 2--20~minutes for baselines of interest for the extraction of the HI signal. The phase of the correlation function can be made small after scaling out an appropriate term, which also causes the coherence time scale of the phase to be longer than the amplitude of the correlation function. We find that our results are insensitive to the input HI power spectrum and therefore they are directly applicable to the analysis of the drift scan data. We also apply our formalism to a set of point sources and statistically homogeneous diffuse correlated foregrounds. We find that point sources decorrelate on a time scale much shorter than the HI signal. This provides a novel mechanism to partially mitigate the foregrounds in a drift scan.
\end{abstract}

\section{Introduction}
The probe of the end of the cosmic dark age remains an outstanding issue in modern cosmology.  From theoretical consideration, we expect the  first luminous objects to appear at a  redshift $z \simeq 30$. The radiation from  these first light sources ionized and heated  the neutral hydrogen (HI)  in their neighbourhood. As the universe evolved,  these ionized regions  grew and merged, resulting in a fully ionized universe by $z\simeq 6$, as suggested by the measurement of Gunn-Peterson troughs of quasars (\citealt{fan06}). Recent Planck  results on CMB temperature and polarization anisotropies fix  the reionization 
epoch    at   $z \simeq  7.7$ (\citealt{planck18}).   The cosmic time  between the  formation of the first light sources ($z \simeq 30$, the era of cosmic dawn) and the universe becoming fully ionized ($z \simeq 6$) is generally referred to  as the  epoch of reionization (EoR). Many important astrophysical  processes during this era, e.g.   the growth and evolution of large scale structures and  the nature of first light sources,  can be  best probed using the hyperfine transition of  HI.  Due to the expansion of the universe,  this line   redshifts   to frequencies  70--200~MHz ($z \simeq 6\hbox{--}20$), which can be detected using meter-wave radio telescopes.

Several existing and upcoming radio telescopes aim to detect the  fluctuating component of this signal, e.g. radio interferometers---Murchison Widefield Array (MWA, \citealt{tingay13_mwasystem}, \citealt{bowman13_mwascience}, \citealt{mwaphaseII_design}), Low Frequency Array (LOFAR, \citealt{vanhaarlem13}), Donald C. Backer Precision Array for Probing the Epoch of Reionization (PAPER, \citealt{parsons14}), Hydrogen Epoch of Reionization Array (HERA, \citealt{deboer17}), Giant Metrewave Radio Telescope (GMRT, \citealt{paciga11}). In addition there are multiple ongoing experiments to detect the global  (sky-averaged)  HI signal from this era---e.g. EDGES, SARAS (\citealt{bowman18}, \citealt{singh18}).

We focus on the fluctuating component of the HI signal in this paper. There are considerable difficulties in the detection of this signal.   Theoretical studies suggest  that the strength of this  signal  is of the order of 10~mK while  the foregrounds are brighter than  100~K (for detailed review see \citealt{furlanetto06}, \citealt{morales10}, \citealt{pritchard12}). These contaminants  include diffuse galactic synchrotron, extragalactic point and extended radio sources, supernova remnants, free free emission, etc. Current experiments can reduce the thermal noise of the system to suitable levels in many hundred   hours of integration. The foregrounds can potentially be mitigated by using the fact that  the  HI signal and its correlations emanate from the three-dimensional   large
scale  structure  at high redshifts. On the  other hand, foreground contamination is dominated by spectrally smooth sources.  This means that even 
if foregrounds can mimic the HI signal on the plane of the sky, the third axis, corresponding to the frequency, can be used to distinguish between the two. All ongoing experiments exploit this spectral distinction  to isolate the  HI signal from foreground contamination (e.g. \citealt{parsons09}, \citealt{parsons12b}).

Using data from ongoing experiments, many  pipelines have been developed to analyze the signal  (\citealt{paul16}, \citealt{paciga11}, \citealt{patil17}, \citealt{beardsley16}, \citealt{choudhuri16}, \citealt{trottchips2016}, \citealt{dillon15}). 
PAPER (\citealt{ali15}) had placed the tightest constraint  on the HI power spectrum but the result has since  been retracted (\citealt{erratum_ali15}).  Their revised upper limit is $(200 ~\rm mK)^2$ at redshift $z = 8.37$ for $k \simeq 0.37 ~\rm Mpc^{-1}$ (\citealt{paper_reanalysis}). 
The current best upper limits on the HI power spectrum are:  $(79.6 ~\rm mK)^2$ , $k\simeq 0.053h \, \rm Mpc^{-1}$,  $z \simeq 10.1$ (LOFAR, \citealt{patil17}) and  $(62.5 ~\rm mK)^2$, $k = 0.2h \, \rm Mpc^{-1}$,  $z \simeq  7$ (MWA, \citealt{barry19_newlim}).
More recently, \cite{bowman18} reported the  detection of  an absorption trough of strength $500 \, \rm mK$ in the  global HI signal in the redshift range $15 < z< 19$. 

Given the weakness of the HI signal, strong foregrounds, and the requirement
of hundreds of hours of integration for detection, one needs extreme stability of the system, precise calibration, and reliable isolation of foregrounds. Drift scans  constitute a   powerful technique  to achieve instrumental stability  during an  observational run. During such a scan the primary beam and other instrumental parameters remain unchanged while the sky  intensity pattern  changes.  Two ongoing interferometers, PAPER and HERA, work predominantly in this mode while the others can also acquire data in this mode. Different variants of drift scans have been proposed in the literature: 
$m$-mode analysis (\cite{m-mode1, m-mode2}, applied to OVRO-LWA data in \cite{eastwood18}), cross-correlation of the HI signal in time (\citealt{paul14}), drift and shift method (\citealt{trott14}) and fringe-rate  method (\cite{fringe-rate}, applied to PAPER data).  \cite{trott14}  provided a framework to estimate the uncertainty in measurement of HI power spectrum based on visibility covariance. Using simulations of visibility covariance, \cite{lanman_corr} have shown that the sample variance can increase up to 20\% and 30\% on the shortest redundant baselines of HERA and MWA respectively.

Owing to changing intensity pattern, it is conceptually
harder to extract   the HI signal from drift scans. As the HI signal is buried 
beneath instrumental noise, it is imperative that correct algorithm 
be applied to retain this sub-dominant component and prevent its loss (e.g. \cite{carina18}). 

In this paper,  we extend the work of \cite{paul14}  to delay space and, additionally, identify the effects of  phase covariance 
and primary beam size. We also apply our formalism to foregrounds by considering  a set of  isotropically-distributed  point sources and statistically
homogeneous correlated diffuse emission.   We 
work in both frequency and   delay space,   the preferred  coordinate for separating  foregrounds from the HI signal (e.g. \citealt{datta10}, \citealt{parsons12b}). 
Our primary aim is to  determine the  correlation time scales of time-ordered visibilities of HI signal in drift scan observations. 
This information can be used to 
establish how the HI signal can be extracted from   drift scans
using correlation of visibilities measured at different times.

 In the next section, we motivate the issue, develop our general  formalism, and apply it to the HI signal in frequency and delay space. We use  primary beams of PAPER, MWA, HERA, and SKA1-Low for our work. We discuss in detail analytic
approximation of   numerical results in the section and Appendix~B.  In section~\ref{sec:fore_drift} we discuss the nature of 
foregrounds and compute the visibility correlation functions for a set of point
sources and diffuse foregrounds. In section~\ref{sec:ana_data_dr}, we elaborate 
on how our formulation can be applied to drift scan data. We discuss many different approaches to the analysis of data including  comparison with earlier attempts.  In the final section,
we summarize our main results.

Throughout this paper we use  spatially-flat  $\Lambda$CDM model  with  $H_0 = 100~\rm h~Km/sec/Mpc$, $\rm h = 0.67$, $\Omega_\Lambda = 0.6911$ (\citealt{planck15}).

\section{HI visibility correlation in Drift scans}\label{sec:h1sig}

The measured visibilities are a function of frequency, baseline, and time. The aim of this section is to determine  the correlation
structure of visibilities in these domains. In particular, our focus is on the correlation structure of visibilities as a function of 
time as the intensity pattern  changes, for a fixed primary beam,   during a drift scan. 

This information allows us to  average  the data in the uv space
with optimal signal-to-noise and prevent
possible HI signal loss. The signal loss could occur if the data is averaged 
over scales larger than the scales of correlation (see e.g. \cite{carina18}). For instance,  the visibilities owing to HI signal are  correlated for baselines
separated by  roughly
the inverse of primary beam, so averaging data over pixels larger than the inverse of primary beam would result in the loss of HI signal. However, if the data is averaged using  pixels much smaller than the correlation scale then it  would 
result in  sub-optimal signal-to-noise. 

In this paper, we determine the time scales over which measured visibilities (for a given baseline, etc.)  are coherent in time and therefore could be averaged in a drift scan to yield optimal signal-to-noise without 
any loss in HI signal.  For this purpose, we derive the correlation function of visibilities, arising from the EoR HI signal,  measured   at two different times in  a drift scan.

A pair of antennas  of a radio interferometer measures the  visibility $V_{\nu}$,  which is related to the sky intensity pattern as (Eq.~2.21 of \citealt{Synthesis_Imaging}):
	\begin{equation}\label{eq:VI}
		V_{\nu}(u_{\nu},v_{\nu},w_{\nu}) = {\int} \frac{dldm}{n} A_{\nu}(l,m)I_{\nu}(l,m)\exp\left[{-2{\pi}i\left( u_{\nu}l + v_{\nu}m + w_{\nu}(n-1) \right)} \right]
	\end{equation}
Here $\nu$ is the observing frequency. ($u_{\nu},v_{\nu}, w_{\nu}$) are the components of the baseline vector between two antennas measured in units of wavelength.  ($l,m,n$) define the direction cosine triplet in the sky with $n = \sqrt{ 1- l^2 - m^2}$. $A_{\nu}(l,m)$ is the primary beam power pattern of an antenna element and $I_{\nu}(l,m)$ is the specific intensity pattern  in the  sky.
We further define vectors $\textbf{u}_{\nu} = (u_{\nu},v_{\nu})$ and  $\boldsymbol{\theta} = (l,m)$. The intensity  pattern owing to  the EoR HI gas distribution  $I_{\nu}(\boldsymbol{\theta})$ can be decomposed in  mean  and fluctuating components as:
	\begin{equation}\label{eq:Icomp}
	I_{\nu}(\boldsymbol{\theta}) = \bar{I}_{\nu} +  \Delta I_{\nu}(\boldsymbol{\theta})
	\end{equation}
As an interferometer measures only fluctuating components of the  signal, we can write: 
	\begin{equation}\label{eq:VdI}
	V_{\nu}(\textbf{u}_{\nu}, w_{\nu}) = {\int} \frac{d^{2}\theta}{n} A_{\nu}(\boldsymbol{\theta}) \Delta I_{\nu}(\boldsymbol{\theta})\exp\left[{-2{\pi}i\left({\textbf{u}_{\nu}}{\cdot}{\boldsymbol{\theta}}+w_{\nu}(n-1)\right)} \right]
	\end{equation}
The HI  inhomogeneities  $\delta_{\rm HI}(\textbf{k})$ arise from various factors such as HI density fluctuations, ionization inhomogeneities,  etc. The  fluctuation in the specific intensity $\Delta I_{\nu}(\boldsymbol{\theta})$ can be related to  the  HI density fluctuations  in the Fourier space,  $\delta_{\rm HI}(\textbf{k})$:
\begin{equation}
	\Delta I_{\nu}(\boldsymbol{\theta})  = \bar{I}_{\nu} {\int} \frac{d^{3}k}{(2\pi)^{3}} \delta_{HI}(\textbf{k}) \exp \left[{i \textbf{k} {\cdot} \textbf{r}}\right] \label{eq:dIdel1}
\end{equation}
Here $\textbf{r}$ is the three-dimensional (comoving) position   vector  and its Fourier conjugate variable is $\textbf{k}$; $k$,  the magnitude of the $\textbf{k}$ vector, is  $k = |\textbf{k}| = \sqrt{k^{2}_{\perp} + k^{2}_{\parallel} } = \sqrt{k^{2}_{\perp 1} + k^{2}_{\perp 2} + k^{2}_{\parallel} }$, where ${\bf k}_\perp$ and $k_\parallel $ are the (comoving) components on the plane of the sky and along the line of sight, respectively. The position vector $\textbf{r}$ can be written in terms of the  line of sight (parallel) and perpendicular components as $\textbf{r} = r_{\nu}{\hat{n}} + r_{\nu} \boldsymbol{\theta}$;  $r_{\nu}$ is the comoving distance.  Eq.~(\ref{eq:dIdel1}) reduces to:
	\begin{equation}\label{eq:dIdel2}
	\Delta I_{\nu}(\boldsymbol{\theta}) = \bar{I}_{\nu} {\int} \frac{d^{3}k}{(2\pi)^{3}} \delta_{\rm HI}(\textbf{k}) \exp\left[{i r_{\nu}\left( k_{\parallel} + \textbf{k}_{\perp} {\cdot} \boldsymbol{\theta}\right)}\right]
	\end{equation}
 As the HI fluctuations are statistically homogeneous, we can define the  HI power spectrum $P_{\rm HI}(k)$ as \footnote{We also assume here that the 
HI signal is statistically isotropic which allows us to write the power spectrum as a function of $|\textbf{k}|$. Statistical isotropy is broken owing to line of sight effects such as redshift space distortion and line-cone anisotropies, which would make the power spectrum  depend on the angle between $\textbf{k}$ and the line of sight. }:
\begin{equation}
	\big\langle \delta_{\rm HI}(\textbf{k}) \delta_{\rm HI}^{*}(\textbf{k}')\big\rangle  = (2\pi)^{3}\delta^{3}(\textbf{k}-\textbf{k}')P_{\rm HI}(k) \label{eq:defpws}
	\end{equation}
In tracking observations,  the  primary beam of the telescope follows  a particular patch of the sky. In a drift scan,  the  sky  pattern  moves with respect to the fixed primary beam. This change  of the sky intensity  with respect to the  fixed phase center  introduces a time dependent  phase $\boldsymbol{\vartheta}(t)$ in  the expression of $\Delta I_{\nu}(\boldsymbol{\theta})$ in Eq.~(\ref{eq:dIdel2}), which gives us  the fluctuating component of the specific intensity as a function of time:
	\begin{equation}\label{eq:dIdel3}
	\Delta I_{\nu}\left(\boldsymbol{\theta},t\right) = \bar{I}_{\nu} {\int} \frac{d^{3}k}{(2\pi)^{3}} \delta_{\rm HI}(\textbf{k}) \exp\left[{i r_{\nu}\left( k_{\parallel} + \textbf{k}_{\perp} {\cdot} \left(\boldsymbol{\theta} - \boldsymbol{\vartheta}(t) \right)\right)}\right]
	\end{equation}
In Eq.~(\ref{eq:VdI}) we use the expression of $\Delta I_{\nu}\left(\boldsymbol{\theta},t\right)$ and expand terms containing $n$ up to first non-zero order 
\footnote{As discussed below, we use primary beams corresponding for many ongoing and future radio telescopes for 
our analysis. For all the cases, this approximation  holds  for the main
lobe of the primary beam, which means, as we show later, that  our main results are unaffected.} 
as ${d^{2}\theta}/{n} \simeq d^{2}\theta$ and $w_{\nu}(n-1) \simeq - \left(l^{2} + m^{2}\right) w_{\nu} /2 = -\theta^{2} w_{\nu} /2 $. This gives us:
\begin{equation}
	V_{\nu}(\textbf{u}_{\nu}, w_{\nu},t) \simeq \bar{I}_{\nu} {\int} \frac{d^{3}k}{(2\pi)^{3}} \delta_{HI}(\textbf{k}) \exp\left[{i r_{\nu} k_{\parallel}}\right] {\int} d^{2}\theta A_{\nu}(\boldsymbol{\theta}) 
	\exp\left[ -2 \pi i \left( \left(\textbf{u}_{\nu} -\frac{r_{\nu}}{2\pi}\textbf{k}_{\perp} \right)\cdot{\boldsymbol{\theta}} + \frac{r_{\nu}}{2\pi}\textbf{k}_{\perp}\cdot{\boldsymbol{\vartheta}(t)} -\frac{1}{2}w_{\nu}\theta^{2}\right) \right] \nonumber
\end{equation}
Next we compute the two-point visibility correlation function between two different frequencies, baselines, and times:
\begin{align}
 \Big\langle V_{\nu}(\textbf{u}_{\nu}, w_{\nu},t)V^{*}_{\nu'}(\textbf{u}'_{\nu'}, w'_{\nu'},t') \Big\rangle \nonumber \\ 
 & \mkern-230mu \simeq \bar{I}_{\nu} \bar{I}_{\nu'} {\int} {\int} \frac{d^{3}k}{(2\pi)^{3}} \frac{d^{3}k'}{(2\pi)^{3}} \big\langle \delta_{\rm HI}(\textbf{k}) \delta_{\rm HI}^{*}(\textbf{k}')\big\rangle \exp\left[{i \left( r_{\nu} k_{\parallel} - r_{\nu'} k'_{\parallel} \right)}\right] {\int} d^{2}\theta A_{\nu}(\boldsymbol{\theta}) {\int} d^{2}\theta' A_{\nu'}(\boldsymbol{\theta'}) \nonumber \\
 & \mkern-215mu \times \exp \left[ -2 \pi i \left( \left(\textbf{u}_{\nu} -\frac{r_{\nu}}{2\pi}\textbf{k}_{\perp} \right)\cdot{\boldsymbol{\theta}} -\left(\textbf{u}'_{\nu'} -\frac{r_{\nu'}}{2\pi}\textbf{k}'_{\perp} \right)\cdot{\boldsymbol{\theta'}} + \frac{r_{\nu}}{2\pi}\textbf{k}_{\perp}\cdot{\boldsymbol{\vartheta}(t)} - \frac{r_{\nu'}}{2\pi}\textbf{k}'_{\perp}\cdot{\boldsymbol{\vartheta'}(t')} -\frac{1}{2} w_{\nu}\theta^{2}  + \frac{1}{2}w'_{\nu'}\theta'^{2} \right) \right] \label{eq:viscor}
\end{align}
  Using Eq.~(\ref{eq:defpws}) in Eq.~(\ref{eq:viscor}) gives the two-point correlation function in terms of the HI power spectrum $P_{\rm HI}(k)$. 
   We first note that the time dependence of Eq.~(\ref{eq:viscor}) 
occurs as the  time difference, $\Delta t$  in   just one term ${\boldsymbol{\vartheta'}(t')} - {\boldsymbol{\vartheta}(t)} = \Delta {\boldsymbol{\vartheta}(\Delta t)}$ which is obtained by dropping the frequency dependence of $r_\nu$. This approximation is discussed in detail in the next subsection.
Eq.~(\ref{eq:angle_expand}) is used to express the time-dependent part of the correlation function 
explicitly in terms of change in the  hour angle $\Delta H$ (for details see
Appendix~A). 
 This gives us:
\begin{align}
	\Big\langle V_{\nu}(\textbf{u}_{\nu}, w_{\nu},t) V^{*}_{\nu'}(\textbf{u}'_{\nu'}, w'_{\nu'},t') \Big\rangle &= 
	\bar{I}_{\nu} \bar{I}_{\nu'} {\int} \frac{d^{3}k}{(2\pi)^{3}} P_{HI}(k) 
	\exp \left[ i k_{\parallel} \left( r_{\nu}  - r_{\nu'} \right) \right] 	\exp \left[i r_{\nu} k_{\perp 1} \cos{\phi} \Delta H \right] \nonumber \\
	& \mkern+80mu \times Q_\nu(\textbf{k}_{\perp}, \textbf{u}_{\nu}, w_{\nu}, \Delta H = 0) Q^*_{\nu'}(\textbf{k}_{\perp}, \textbf{u}'_{\nu'}, w'_{\nu'}, \Delta H)	\label{eq:viscorapprox2}
\end{align}
Here   $\phi$ is the latitude of the telescope and the Fourier beam (or 2D $Q$-integral) is defined as:
\begin{align}
	 Q_\nu(\textbf{k}_{\perp}, \textbf{u}_{\nu}, w_{\nu}, \Delta H) &= {\int} d^{2}\theta A_{\nu}(\boldsymbol{\theta}) 
	\exp\left[ -2 \pi i \left( \boldsymbol{x}_u \cdot {\boldsymbol{\theta}} -\frac{1}{2} y \theta^{2}\right) \right] \label{eq:viscorapprox3} \\
  \text{with}	\qquad \qquad
	x_u &= u_{\nu} - \frac{r_{\nu}}{2\pi} \left(k_{\perp 1} + k_{\perp 2} \sin{\phi} \Delta H \right) \label{eq:def_x_u}\\
	x_v &= v_{\nu} - \frac{r_{\nu}}{2\pi} \left(k_{\perp 2} - k_{\perp 1} \sin{\phi} \Delta H \right) \label{eq:def_x_v}\\
	y   &= w_{\nu} + \frac{r_{\nu}}{2\pi} k_{\perp 1} \cos{\phi} \Delta H \label{eq:def_y}
\end{align} 

In this paper we consider only the zenith drift
scan. Non-zenith drift scans can be treated  by replacing $\phi$ with $\phi+\chi$, where $\chi$ is the angle between the latitude of the zenith and the phase center of the observed  field (for details see Appendix~A in \citealt{paul14}). This doesn't impact our main results.
Eq.~(\ref{eq:viscorapprox2}) can be numerically solved for a given primary beam pattern $A_\nu(\boldsymbol{\theta})$. We next discuss the visibility correlation in delay space, the preferred coordinate  for analysing the data.

\subsection{Visibility Correlation in Delay Space} \label{sec:delay_space}

To compute  the HI visibility correlation function in  delay space ($\tau$) we  define:
\begin{align}\label{eq:defdelay}
 V_{\tau}(\textbf{u}_{0}, w_{0},t) = \int_{\nu_0 - B/2}^{\nu_0 + B/2} d\nu V_{\nu}(\textbf{u}_{\nu}, w_{\nu},t) \exp\left[{2 \pi i \tau \nu}\right]
\end{align}
Throughout this paper the subscript `0' under any variable denotes the value of that variable at the central frequency. Throughout this paper, we use:  $\nu_0 \simeq 154~\rm{MHz}$ and  bandpass $B \simeq 10~\rm{MHz}$. 
Its cross-correlation in delay space can be expressed as:
\begin{align}
 \Big\langle V_{\tau}(\textbf{u}_{0}, w_{0},t)V^{*}_{\tau}(\textbf{u}'_{0}, w'_{0},t') \Big\rangle = \iint_{\nu_0 - B/2}^{\nu_0 + B/2} d\nu d\nu' \Big\langle V_{\nu}(\textbf{u}_{\nu}, w_{\nu},t)V^{*}_{\nu'}(\textbf{u}'_{\nu'}, w'_{\nu'},t') \Big\rangle e^{-2 \pi i \tau \Delta \nu }
\label{eq:visdelaysp}
\end{align}
Eq.~(\ref{eq:visdelaysp}) can be reduced to a  more tractable form by making 
appropriate approximations. We expand frequency-dependent variables 
in exponents around $\nu_0$ up to the first order. Thus $\left( r_{\nu}  - r_{\nu'} \right) \simeq -\dot{r}_{0} \Delta \nu$, denoting  $\left( dr_{\nu}/d\nu \right)_{\nu_0} = \dot{r}_{0}$, $\nu' - \nu = \Delta \nu $. To the same order, the 
approximation made following Eq.~(\ref{eq:viscorapprox2}) is also valid.  We further approximate  $\textbf{u}_{\nu} \simeq \textbf{u}_{0}$ and drop the weak frequency dependence of the mean specific intensity and primary beam within the observing band-width $B$.  We discuss the impact of these
approximations in section~\ref{sec:app_inp}.
This gives us:
\begin{align} \label{eq:viscorapprox4}
 \Big\langle V_{\tau}(\textbf{u}_{0}, w_{0},t)V^{*}_{\tau}(\textbf{u}'_{0}, w'_{0},t') \Big\rangle =&
\bar{I}^{2}_{0} {\int} \frac{d^{3}k}{(2\pi)^{3}} P_{\rm HI}(k) \exp\left[{i r_{0} k_{\perp 1} \cos{\phi} \Delta H}\right]
\left(  \iint_{\nu_0 - B/2}^{\nu_0 + B/2} d\nu d\nu'  \exp\left[-i \Delta\nu \left(k_{\parallel} \dot{r}_{0}  +2 \pi  \tau  \right)\right] \right)\nonumber \\
& \times Q_{\nu_0}(\textbf{k}_{\perp},\textbf{u}_{0},w_{0}, \Delta H = 0) Q_{\nu_0}^{*}(\textbf{k}_{\perp},\textbf{u}'_{0},w'_{0},\Delta H) 
\end{align}
The integrals over $\nu$ and $\nu'$ can be solved in two ways. They can be solved by changing the variables from ($\nu, \nu'$) to ($x, y$). $x = \nu'-\nu = \Delta \nu$ and $y = \left( \nu'+\nu \right)/2$. They  can also be solved by separating $\Delta \nu = \nu'-\nu$ and integrating over $\nu$ and $\nu'$ individually.  The resulting function peaks sharply at $\tau = - \dot{r}_{0}k_{\parallel}/(2\pi)  $. The major contribution to the integral in Eq.~($\ref{eq:viscorapprox4}$) occurs  when $k_{\parallel} = - 2\pi \tau/\dot{r}_{0}$, which gives us  the well-known correlation scale along the line-of-sight direction (e.g. \citealt{paul16}). We use the $\delta$-function approximation for  frequency integrals:
\begin{align}
\iint_{\nu_0 - B/2}^{\nu_0 + B/2} d\nu d\nu'  \exp\left[-i \Delta\nu \left(k_{\parallel} \dot{r}_{0}  +2 \pi  \tau  \right)\right] =
B^{2} \sinc^{2}\left[\pi B \left( \tau + \frac{\dot{r}_{0}}{2\pi} k_{\parallel} \right) \right ] \simeq \frac{2\pi B}{|\dot{r}_{0}|} \delta \left( k_{\parallel} - \frac{2\pi \tau}{|\dot{r}_{0}|} \right) \label{eq:delfnapprox}
\end{align}
This approximation preserves the area under the curve. We note that the delta function approximation used in Eq.~(\ref{eq:delfnapprox}) could break down
  if $B$ is small. For $B = 10 \, \rm MHz$, we use in the paper, it is an excellent assumption.
  For a much smaller $B$, the sinc function in the equation can be directly integrated without making any difference to our main results. We denote $\dot{r}_{0} = -|\dot{r}_{0}|$ because the comoving distance decreases with increasing frequency. Using this  in Eq.~(\ref{eq:viscorapprox4}) we find, with $k_{\parallel} = 2\pi \tau/|\dot{r}_{0}|$:
\begin{align} \label{eq:viscorapprox5}
 \Big\langle V_{\tau}(\textbf{u}_{0}, w_{0},t)V^{*}_{\tau}(\textbf{u}'_{0}, w'_{0},t') \Big\rangle \simeq &
\bar{I}^{2}_{0} \frac{B}{|\dot{r}_{0}|}{\int} \frac{d^{2}k_{\perp}}{(2\pi)^{2}} P_{\rm HI}(k) \exp \left[ {i r_{0} k_{\perp 1} \cos{\phi} \Delta H} \right] \nonumber \\
& \mkern-20mu \times Q_{\nu_0}(\textbf{k}_{\perp},\textbf{u}_{0},w_{0}, \Delta H = 0) Q_{\nu_0}^{*}(\textbf{k}_{\perp},\textbf{u}'_{0},w'_{0},\Delta H)
\end{align}
Here $k=\sqrt{k_{\perp 1}^2+k_{\perp 2}^2 + (2\pi\tau/|\dot{r}_{0}|)^2}$. Eq.~(\ref{eq:viscorapprox5}) generalizes the results of \cite{paul14}  to delay space and also accounts for the impact of the  $w$-term.
To further simplify Eq.~(\ref{eq:viscorapprox5})  we need an expression for the primary beam pattern. We consider four radio interferometers in 
our analysis. 

MWA: MWA has square-shaped antennas called tiles. Each tile consists of 16 dipoles placed on a mesh and arranged in a 4x4 grid at spacing of roughly 1.1~meters. 
Effective area of a tile $A_{\rm eff} = 21.5 ~\rm m^2 $ at 150~MHz (\citealt{tingay13_mwasystem}). 
The square of the absolute value of the 2D Fourier transform of the antenna shape gives the antenna power response. For MWA $A_{\nu}(l,m) = \sinc^{2}(\pi L_{\nu} l)\sinc^{2}(\pi L_{\nu} m)$. Here $L_{\nu} = L \left( \nu/\nu_{0} \right)$; $L \left(\equiv \sqrt{A_{\rm eff}}/\lambda_0 \simeq 2.4 \right)$ is the length of the square tile in units of central wavelength ($\lambda_0 \simeq 1.95 \rm m$).   Therefore, the 2D primary beam response $ A_{\nu}(l,m)$  
can be represented as a product of two independent 1D patterns;  $A_{\nu}(l,m) = A_{\nu}(l) A_{\nu}(m)$. 

PAPER, HERA and SKA1-Low: Individual element in PAPER, HERA,  and SKA1-Low   correspond to dishes  of diameter 2~meters,  14~meters,  and 35~meters, respectively.  The beam pattern at a 
frequency $\nu$ can be expressed as: 
$A_{\nu} = 4 |j_1(\pi d_\nu \sqrt{l^2+m^2})/(\pi d_\nu \sqrt{l^2+m^2}))|^2$, where $j_1(x)$ is the spherical Bessel function and  $d_\nu$ is the diameter of the dish in the units of wavelength. Unlike MWA, this primary beam pattern is not separable in $l$ and $m$. Or the double integral over angles in Eq.~(\ref{eq:viscorapprox3}) cannot be expressed as a product 
of two  separate  integrals over $l$ and $m$.  We do not consider LOFAR in our analysis as its core primary beam, suitable for EoR studies, is close  to SKA1-Low
\footnote{ http://old.astron.nl/radio-observatory/astronomers/lofar-imaging-capabilities-sensitivity/lofar-imaging-capabilities/lofa}.
For MWA and SKA1-Low: $\phi = -26.7^\circ$ and for  HERA and PAPER: $\phi = -30.7^\circ$.

\begin{figure}
	\centering
	\begin{minipage}{0.49\textwidth}
		\centering
		\includegraphics[width=1.0\linewidth]{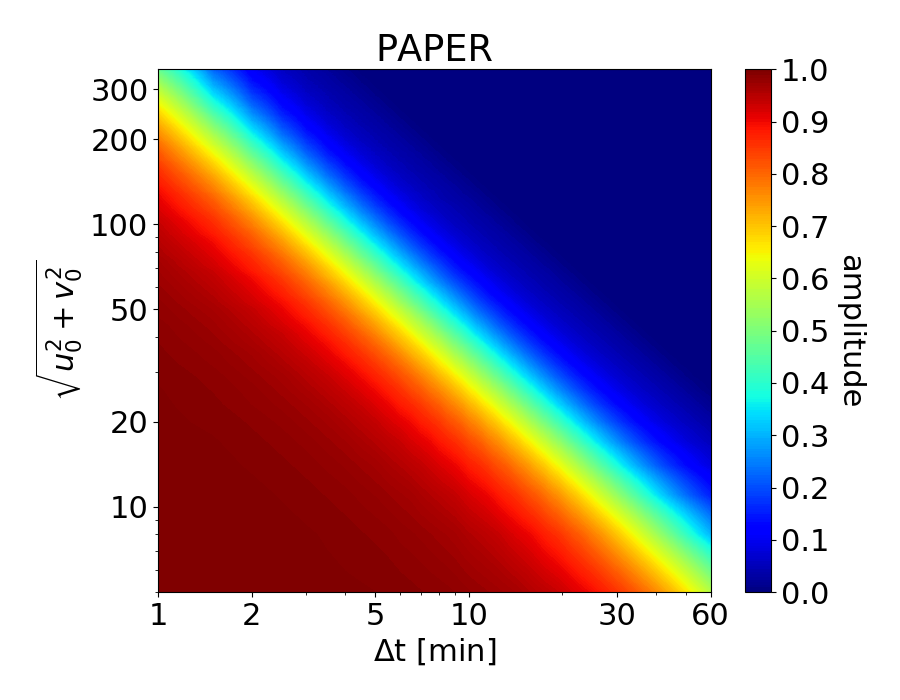}
	\end{minipage}\hfill
	\centering
	\begin{minipage}{0.49\textwidth}
		\centering
		\includegraphics[width=1.0\linewidth]{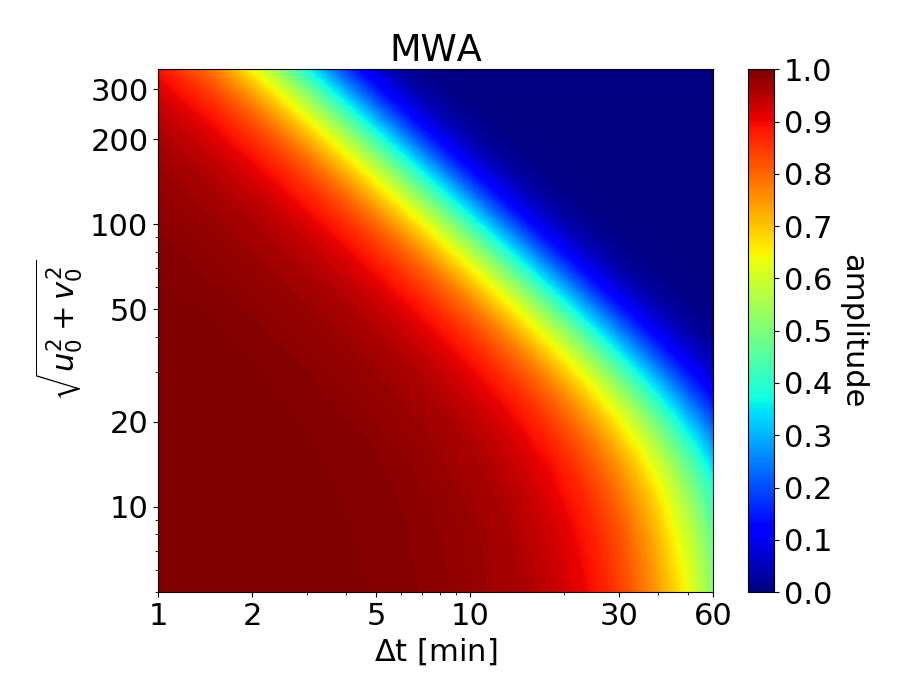}
	\end{minipage}\hfill
\centering
	\begin{minipage}{0.49\textwidth}
		\centering
		\includegraphics[width=1.0\linewidth]{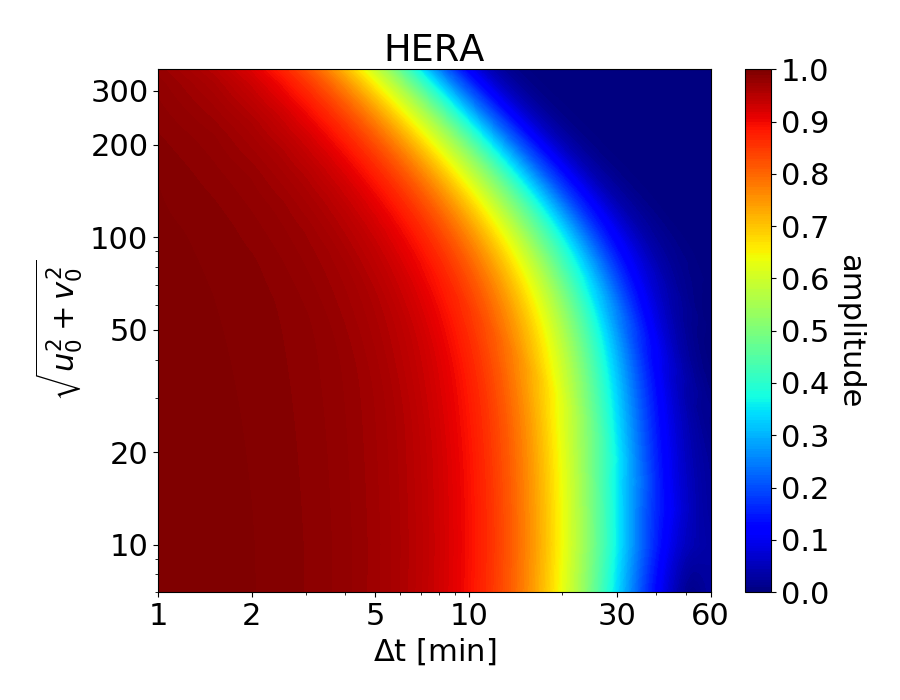}
	\end{minipage}\hfill
	\centering
	\begin{minipage}{0.49\textwidth}
		\centering
		\includegraphics[width=1.0\linewidth]{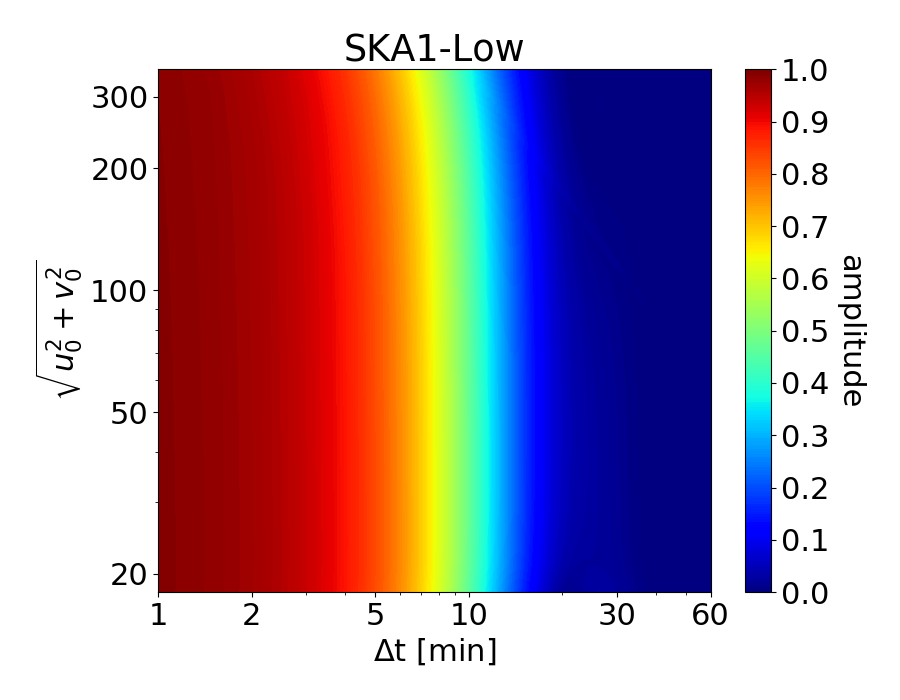}
	\end{minipage}\hfill
\caption{ 
The  figure displays  the amplitude of the visibility correlation function as a function of $\Delta t$, normalized to unity for $\Delta t=0$. The  quantity plotted in the figure  is $ \langle V_{\tau}(\textbf{u}_{0}, w_{0},t)V^{*}_{\tau}(\textbf{u}_{0}, w_{0},t') \rangle  / \langle V_{\tau}(\textbf{u}_{0}, w_{0},t)V^{*}_{\tau}(\textbf{u}_{0}, w_{0},t) \rangle$ as a function of   baseline length  $|\textbf{u}_{0}| = \sqrt{u_0^2+v_0^2}$ and $\Delta t = t'-t$, for  $u_0=v_0$,  $w_0 = 0$, and $\tau = 0$.  
The amplitude of the correlation function decorrelates mainly due to the rotation of the intensity pattern. 
However the impact of the  traversal of the intensity pattern becomes important for smaller primary  beams on  small baselines. 
As seen in the figure, for  all baselines for PAPER and   large baselines for MWA, HERA, and SKA1-Low, the decorrelation time scales are  proportional to  $1/|\textbf{u}_0|$ and  $1/\sqrt{\Omega}$. This effect is discussed  in subsection~\ref{sec:Qinte} (point (b)). On smaller  baselines in MWA, HERA, and SKA1-Low panels, the traversal of the intensity pattern starts dominating the decorrelation. This effect is discussed  in subsection~\ref{sec:Qinte} (point (a)).
}
\label{fig:HIcorr2}
\end{figure}

\begin{figure}
	\centering
	\begin{minipage}{0.46\textwidth}
		\centering
		\includegraphics[width=1.0\linewidth]{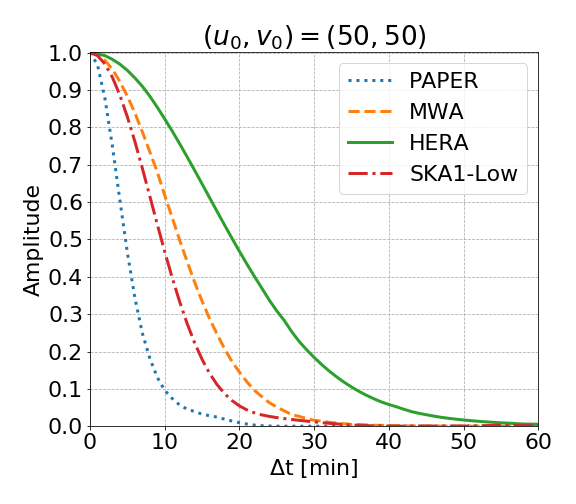}
	\end{minipage}\hfill
	\centering
	\begin{minipage}{0.53\textwidth}
		\centering
		\includegraphics[width=1.0\linewidth]{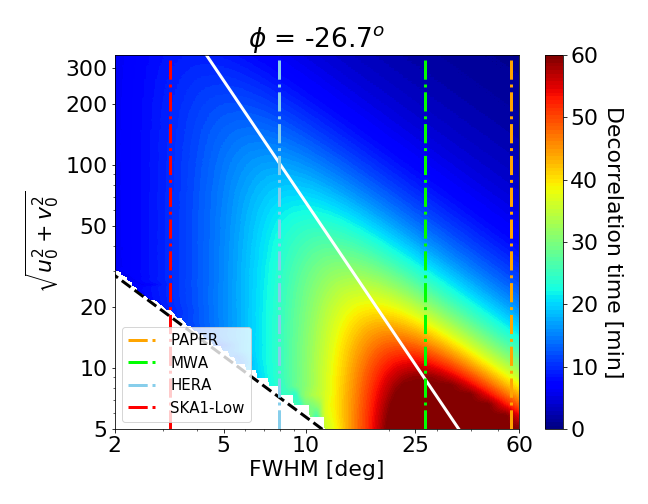}
	\end{minipage}\hfill
        \caption{  
        Left Panel: The amplitude of the visibility correlation function is shown as a function of $\Delta t$ for a fixed baseline for different
          primary beams. Right Panel: The isocontours of the decorrelation time
          are shown in the primary beam--baseline plane; the decorrelation time is defined as  $\Delta t$  such that the amplitude of  correlation function falls to half its value as compared to $\Delta t = 0$. 
          The Figure assumes Gaussian beams (Eq.~\ref{eq:Gauss_beam}) with  ${\rm FWHM} =  2\sqrt{\log_e(2)\Omega_{0g}}$. The region on the left bottom is excluded because the shortest baseline
          $\sqrt{u_0^2 + v_0^2} = d_0$, where $d_0$ is  the primary element of the telescope in units of the central wavelength, $\lambda_0$. There could be minor differences between this figure and 
          Figure~\ref{fig:HIcorr2} because we use  a fixed telescope latitude  $\phi = -26.7$ for all  primary beams. The primary beams of the four interferometers studied in this paper  are marked on the figure. The White line demarcates the  regions dominated by rotation (above the line)  and traversal of intensity pattern (for further discussion see the text). 
}
\label{fig:HIcorr_cov}
\end{figure}

In Figure~\ref{fig:HIcorr2} we show the  amplitude of the correlation function (Eq.~(\ref{eq:viscorapprox5})),  normalized to unity for $\Delta t=0$, as a function of the time difference, $\Delta t \equiv t'-t$ in a drift scan.  In the Figure, we use the HI power spectrum $P_{\rm HI}(k)$ given by the simulation of \cite{furlanetto06}; we discuss the dependence of our results on
the input power spectrum below  in subsection~\ref{sec:pk_effect} . 
The figure displays numerical  results for different primary beams 
as a function of  baselines length $|\textbf{u}_{0}| = \sqrt{u_0^2+v_0^2}$, for  $w_0 = 0$ and  $\tau = 0$. 
Our numerical results further show  that   the visibility correlation function
in time domain is nearly  independent of $\tau$. This 
is discussed and justified in Appendix~B using analytic approximations.
 Figure~\ref{fig:HIcorr_cov} complements
Figure~\ref{fig:HIcorr2} and allows us to study the change in decorrelation time when the primary beam is changed for a fixed baseline; it will be discussed
in detail in the next sub-section.

To get analytic insights into the nature of numerical  results displayed in Figures~\ref{fig:HIcorr2} and~\ref{fig:HIcorr_cov}, we consider a  separable and  symmetric Gaussian beam.

\subsubsection{Fourier Beam  and HI Correlation with Gaussian Beam}\label{sec:Qinte}
The Fourier Beam introduced in Eq.~(\ref{eq:viscorapprox3}) is the response of the primary beam in  the Fourier domain.
It has two useful properties which makes the  computation of the  Fourier beam easier. 
If the primary beam is separable, $A_{\nu}(l,m) = A_{\nu}(l) A_{\nu}(m)$, then the Fourier beam is also separable, 
$Q_\nu(\textbf{u}_{\nu}) = Q^1_\nu(u_{\nu}) Q^2_\nu(v_{\nu})$. 
And  if the 1D primary beam response, $A_{\nu}(l)$, is an  even function then  the 1D Fourier beam, $Q^1_\nu(u_\nu)$, satisfies the  following relations.
\begin{align}\label{eq:Qproperties}
	Q^1_\nu(-x_u,y) &= Q^1_\nu(x_u,y) \nonumber\\
	Q^1_\nu(x_u,-y) &= Q^{1*}_\nu(x_u,y) 
\end{align}
The  expressions above  are also valid for $Q^2_\nu(v_\nu)$. 
This  shows that it is sufficient to calculate Fourier beam for only $x_u, y \geq 0$.
The variables $x_u, x_v, \text{ and } y$ are defined in Eqs.~(\ref{eq:def_x_u})--(\ref{eq:def_y}). $x_u$ and $x_v$ determine the correlation scales in the neighbourhood of  the Fourier mode, $2\pi \textbf{u}_0 / r_0$, at which the $Q$-integral receives maximum contribution. The variable $y$ can be viewed as an effective $w$-term.
We note that when $y$ 
is small $Q^1_\nu(x_u,y)$ is large  but falls very rapidly along $x_u$. For larger values of $y$, $Q^1_\nu(x_u,y)$ is smaller and goes to zero slowly along $x_u$. This behaviour can be understood as follows:  the effective  beam size shrinks for larger value of $w$-term, resulting in a decrease in signal strength but an increase in the correlation scale (e.g. \citealt{paul16,cornwell08}).

The discussion also applies to 2D Fourier beams. The 2D Fourier beam is a function of Fourier coordinates $x_u, x_v$ and parameter $y$. The point $(x_u, x_v) = (0,0)$ receives the  maximum contribution and picks out  Fourier modes, $k_{\perp1}, k_{\perp2}$. 
Large beams have smaller Fourier beams e.g. for PAPER the Fourier beam is the smallest of all the cases we consider. The width of the Fourier beam decides the range of   correlation scales of the HI signal. This range  is roughly on the order of $2/\sqrt{\Omega} \simeq 2 d$  where  $\Omega$ is the primary beam solid angle  and 
$d$ is the antenna size in the  units of wavelength.
The amplitude of the  Fourier beam is more sensitive to $y$ if the beam is larger (PAPER, MWA).

To gain further analytic insights into the HI correlation function,  we use a Gaussian primary beam 
in our  formalism to compute  the Fourier beam. 
For illustration, we choose Gaussian primary beam of solid angle $\Omega_{0g}$  at $\nu_0 =154.24 \, \rm MHz$ ($\Omega_{0g} = 0.25/L^2$ roughly matches the MWA primary beam). This gives us:
\begin{align}\label{eq:Gauss_beam}
	A_{\nu_0}(l,m) = e^{-(l^2 + m^2)/\Omega_{0g}} 
\end{align}
To compute the Fourier response of a Gaussian beam analytically,  we extend the limits of the integral from $[-1,1]$ to $[-\infty,\infty]$, which is a valid 
procedure as the integrand falls rapidly outside the support of the  primary beam.  Using Eq.~(\ref{eq:viscorapprox3}),  we obtain:
\begin{align}
Q_{\nu_0}(\textbf{k}_{\perp}, \textbf{u}_{0}, w_{0}, \Delta H) = \frac{\pi \Omega_{0g}} {1 - i \pi y \Omega_{0g} } \exp \left[ - \frac{\pi^2 \Omega_{0g} (x_u^2 + x_v^2) } {1 - i \pi y \Omega_{0g}} \right] \label{eq:qint1}
\end{align}

We assume $ {\bf u_0 } = {\bf u'_0 } $ and  ${\bf k_\perp } = (2\pi/r_0) {\bf u_0 }$ to study the time behaviour of the correlation function relevant 
in a drift scan. 
The time-dependent part of the visibility correlation function is  determined  by the product of two Fourier beams separated by drift time $\Delta H$ in Eq.~(\ref{eq:viscorapprox5}).
For Gaussian beam this product is:
\begin{align}
Q_{\nu_0}(\Delta H = 0) Q_{\nu_0}^{*}(\Delta H) &=
\frac{ (\pi \Omega_{0g} )^2 }{ ( 1 - i \pi \Omega_{0g} w_0 ) (1 + i \pi \Omega_{0g} y) } 
\exp \left[ - \frac{\pi^2 \Omega_{0g} |{\bf u_0}|^2 \sin^2{\phi} \Delta H^2  } {1 - i \pi y \Omega_{0g}} \right] 
\label{eq:qprod1}
\end{align}
where only the dependence on the  time variable is retained  in the LHS  for brevity. As discussed above,   $y =( w'_{0} + u_0 \cos{\phi} \Delta H)$ acts  as an effective $w$-term. For a zenith drift scan we study in this paper, the $w$-term is small, so we   put $w_0 = w'_0 = 0$. We find the  amplitude  of the product of  the Fourier beams to be:
\begin{align}
|Q_{\nu_0}(\Delta H = 0) Q_{\nu_0}^{*}(\Delta H)| &=
\frac{ (\pi \Omega_{0g} )^2 }{ \sqrt{ (1 + \pi^2 \Omega^2_{0g} u^2_0 \cos^2{\phi} \Delta H^2) } } 
\exp \left[ - \frac{\pi^2 \Omega_{0g} |{\bf u_0}|^2 \sin^2{\phi} \Delta H^2  } {1 + \pi^2 \Omega^2_{0g} u^2_0 \cos^2{\phi} \Delta H^2 } \right]
\label{eq:viscorgau}
\end{align}
Eq.~(\ref{eq:viscorgau}), along with Eqs.~(\ref{eq:viscorapprox5}) and (\ref{eq:qprod1}),  allows us to read off several salient  features of the visibility correlation function in a drift scan.

  Due to the rotation of the  earth on its axis, the  sources in the sky move  with respect to 
the fixed phase center ($l = 0$, $m=0$) of a telescope located at latitude $\phi$. The changing  intensity pattern  is a  combination of two motions: rotation around a fixed phase center  and the  east-west translation of the  pattern with respect to the fixed phase center (Eq.~(\ref{eq:dldmdn})). In Fourier space, the rotation  causes a time-dependent mixing of Fourier modes in the plane of the sky,  while the translation introduces a new time-dependent phase which is proportional to $k_{\perp 1}$,  the component of  the Fourier mode in  the east-west direction (Eq.~\ref{eq:angle_expand})). In addition to these two effects, which are linear in the angle, we also  retain  a second order term  which becomes important for large beams (Eqs.~(\ref{eq:dldmdn}) and~\ref{eq:angle_expand})). The impact of each of these effects
  on the visibility correlation function is discussed  next:

\begin{itemize}
\item[(a)] {\it Traversal time of coherence scale}: The phase term proportional 
to $\exp(i r_{0} k_{\perp 1} \cos{\phi} \Delta H)$ in Eq.~(\ref{eq:viscorapprox5}) represents this effect. $\Delta H \simeq 1/(r_{0} k_{\perp 1} \cos{\phi})$ is the time over which a coherent feature of linear  size $1/k_{\perp 1}$ is traversed in the east-west direction. As $r_0 k_{\perp 1} \simeq 2\pi u_0$, $\Delta H \simeq 1/(2\pi u_0 \cos{\phi})$ appears to give a  rough estimate of the  time over which  the decorrelation occurs  for a given $u_0$, the east-west component of the  baseline. However, it doesn't give a reasonable estimate for the decorrelation time 
scale of the amplitude of the correlation function as Eq.~(\ref{eq:viscorapprox5}) can be multiplied and divided by $\exp(i2\pi u_0 \cos{\phi} \Delta H)$
which allows us to absorb  the fastest changing term as the  phase term of the correlation function. The correlation time scale of the amplitude of the correlation function
depends on the slow phase $\exp(i (r_0 k_{\perp 1}-2\pi u_0) \cos{\phi} \Delta H)$ whose contribution to the visibility correlation  is determined by the primary beam as we discuss below. 
\item[(b)] {\it Rotation of intensity pattern}: 
This effect is captured by
the numerator in the Gaussian in Eq.~(\ref{eq:viscorgau}), which shows that  the decorrelation  owing to the  rotation of the intensity pattern   is proportional
to $1/(\Omega_{0g}^{1/2} |{\bf u_0}| |\sin{\phi}|)$. This effect, unlike (a),  depends  the magnitude of the baseline and not its east-west component. 
 Eqs.~(\ref{eq:viscorapprox3})--(\ref{eq:def_x_v}), along with Eq.~(\ref{eq:dldmdn}) and Eq.~(\ref{eq:angle_expand}),
allow us to understand this effect. When visibilities at two times
are correlated for a given baseline, they respond to different Fourier modes
of the HI power spectrum owing to the rotation of intensity pattern in a drift scan (Eqs.~(\ref{eq:viscorapprox3})--(\ref{eq:def_x_v})). 
The extent of correlation of visibilities which get contribution from different Fourier modes 
depends on the primary beam: the smaller the primary beam the larger is the range of Fourier modes that contribute to the correlation. 
Therefore, the decorrelation time   is proportional to $\Omega_{0g}^{-1/2}$. 
\item[(c)] {\it Large field of view}: 
The terms proportional to $\Omega^2_{0g}$ in
Eq.~(\ref{eq:viscorgau}) (or more generally the terms proportional to $y$ in Eq.~(\ref{eq:qint1})) are responsible for this effect. These terms correspond to an effective $w$-term, a part of which arises from  $w_0$ and the remaining is the higher-order  time-dependent phase  in a drift scan. This effect is important when the primary beam or  $w_0$ is large. \footnote{Throughout our analysis we assume $w_0 =0$ and we only consider the impact of the time-dependent term. Our assumption would be valid for a  zenith drift scan, which we assume,   for a near-coplanar  interferometric  array. Coplanarity is generally a good assumption as our focus for the detection of the HI signal is short baselines, e.g. for MWA $w_0\ll |\textbf{u}|$ for a zenith scan. We can gauge  the quantitative impact of non-zero $w_0$ using Eq.~(\ref{eq:qprod1}). The main effect of non-zero $w_0$ is to yield a smaller effective  primary beam (\cite{paul16,paul14,cornwell08}) and to introduce additional phase in the visibility correlation function (Eq.~(\ref{eq:phase_gua})). } 
\end{itemize}
We next  discuss the relative importance of (a), (b), and (c) in understanding 
Figures~\ref{fig:HIcorr2} and~\ref{fig:HIcorr_cov}. We first note that (c) doesn't play an important role 
in explaining qualitative features seen in the Figures. Its impact is only mildly important 
for PAPER at the smallest baselines we consider. 

For PAPER, the decorrelation 
time in the Figure scales   linearly as the inverse of the length of the baseline $1/|{\bf u_0}|$. Figure~\ref{fig:HIcorr2} shows only the case $u_0=v_0$. We have checked that the behaviour seen in the figure  is nearly independent of the individual components of the baseline. Also a comparison of decorrelation times between 
PAPER and MWA shows that  the decorrelation times scale as $\Omega_{0g}^{-1/2}$  for baseline $ |{\bf u_0}|  \ga 25$. 
A comparison of these two cases with large baselines $|{\bf u_0}| \ga 150$  for HERA and SKA1-Low also shows the same scaling with the primary beam.  This means that (b) is the dominant decorrelation mechanism in all these  cases. 

For short baselines for  MWA, HERA,  and SKA1-Low the behaviour is markedly different. If (b) alone determined the decorrelation in these cases, the decorrelation time  would be longer as the primary beam is smaller in these two cases, but this behaviour is seen only for longer baselines. Therefore, (a) plays an important role in these cases. 
For large primary beams, (a) is unimportant because the slow phase discussed above is closer to zero, as it gets contribution from a small range of Fourier modes.  However, for narrower  primary beams, this term gets contribution from 
a larger range of Fourier modes which results in cancellation when integration over $k_{\perp 1}$ is carried out. This results in a reduction of correlation time scale. This effect is more dominant for smaller baselines for the following reason:  for a given $u_0$, the range of Fourier modes that contribute to the
  visibility correlation function is $\Delta k_{\perp 1} \simeq 1/(r_0 \Omega_{0g}^{1/2})$ (i.e. size of the Fourier beam) centered around $k_{\perp 1} = 2\pi u_0/r_0$ (e.g. Eqs.~(\ref{eq:viscorapprox3})--(\ref{eq:def_x_v})). It should be noted that $\Delta k_{\perp 1}$ is only determined by the size of the primary beam while $k_{\perp 1}$  scales with the east-west component of the baseline. This implies that  for long baselines, $k_{\perp 1} \gg \Delta k_{\perp 1}$. In this case, the visibility correlation function is dominated by the contribution of a single Fourier mode, which suppresses the impact of possible cancellation that occurs owing to the mixing of Fourier modes, diminishing the impact of (a) for long baselines.  However, when   $\Delta k_{\perp 1} \simeq k_{\perp 1}$, the effect becomes important and it  determines the 
decorrelation time scale for shorter baselines. 

For small baselines and narrower primary beams, both (a) and (b) play an important role so it is worthwhile to investigate the dependence of the decorrelation time on the components of baselines (Figure~\ref{fig:HIcorr2} assumes $u_0=v_0$). We have checked many different combinations of $u_0$ and $v_0$ and find 
that  the qualitative features of Figure~\ref{fig:HIcorr2} are  largely determined by the 
the length of the baseline.  But, as discussed below,
the  phase of the correlation function is dominated  by the east-west component of the baseline.

The correlation structure in the primary-beam--$\Delta t$--baseline space
  is further explored in Figure~\ref{fig:HIcorr_cov}. 
  In the left panel, we show the amplitude of the  correlation function as a function of $\Delta t$ for a fixed
  baseline for different primary beams. 
  The right panel shows the
  isocontours of the decorrelation time  in the primary beam--baseline plane;
  the decorrelation time is defined as 
   the time difference $\Delta t$
  at which the amplitude of the correlation function falls to half its
  value at  $\Delta t=0$.  For each baseline,
  the decorrelation time reaches a maximum value as a function of the  primary beam. Our formalism allows us to understand this general behaviour: for smaller primary beam, the Fourier beam is large which causes decorrelation owing to mode-mixing in the transverse  motion  of the intensity pattern (point (a)). For larger primary beam, the rotation of intensity pattern is responsible for the decorrelation  (point (b)). The decorrelation time scales inversely with the  baseline length and could reach an hour for the shortest baselines and large primary beams, in agreement with Figure~\ref{fig:HIcorr2}.  A notable feature of Figure~\ref{fig:HIcorr_cov} is the alignment of the isocontours of decorrelation time. Its shape is determined by
  the interplay of decorrelation owing to the rotation  and the traversal  of
  the intensity pattern and
  can be derived analytically.

  For large primary beams, the decorrelation time is $\simeq  1/( |{\bf u_0}| \Omega_{0g}^{1/2} |\sin{\phi}| ) $ (point (b), (Eq.~(\ref{eq:viscorgau})); 
  the decorrelation profile  for large
  primary beams is seen to follow this function. For small primary beams, the decorrelation time
  is $\simeq \Omega_{0g}^{1/2}/\cos{\phi}$, nearly independent of the length of the baseline (point (a)). 
  Equating these two expressions gives us: $\Omega_{0g} |\tan{\phi}|~|{\bf u_0}| \simeq 1$. This relation is shown in Figure~\ref{fig:HIcorr_cov} (White line) and it separates the regions dominated by decorrelation owing to
  the rotation (above the White line) from the regions in which the translation
  plays the dominant role. Figure~\ref{fig:HIcorr_cov}
  shows the White line adequately captures the essential physics of the separation of the two regions. We note that the large field of view  (point (c) above) does not play an important role in our study because of the range of telescope latitudes
  we consider, which is motivated by the location of radio interferometers studied here. For $\phi \simeq  90^\circ$,  both translation and large field of view effects  are negligible  while, for $\phi \simeq 0$, the impact of rotation
  is negligible while translation and wide field of view effects dominate (Eq.~(\ref{eq:viscorgau})).

\begin{figure}
	\centering
	\begin{minipage}{0.49\textwidth}
		\centering
		\includegraphics[width=1.0\linewidth]{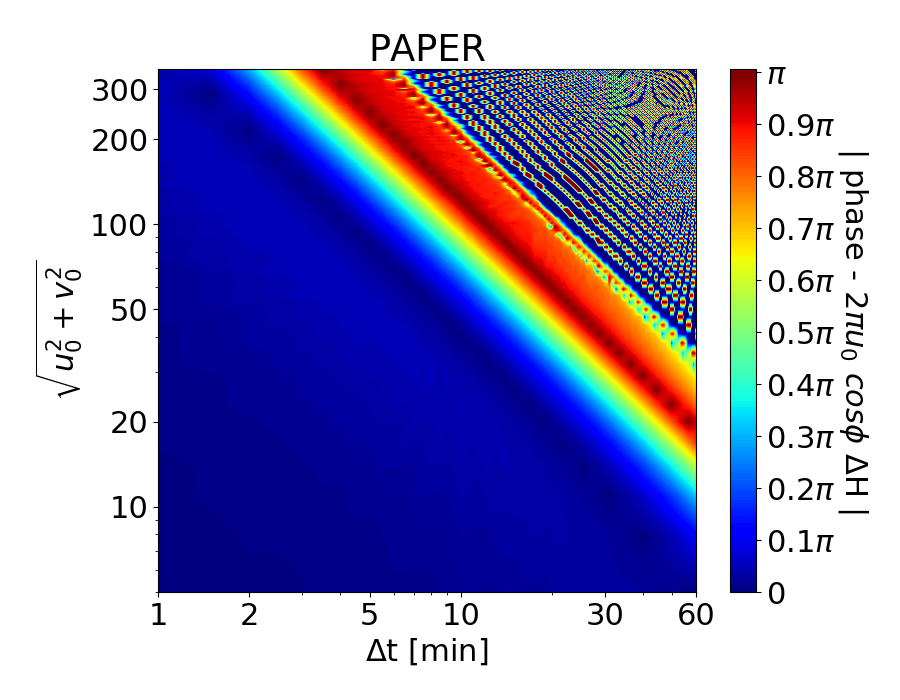}
	\end{minipage}\hfill
	\centering
	\begin{minipage}{0.49\textwidth}
		\centering
		\includegraphics[width=1.0\linewidth]{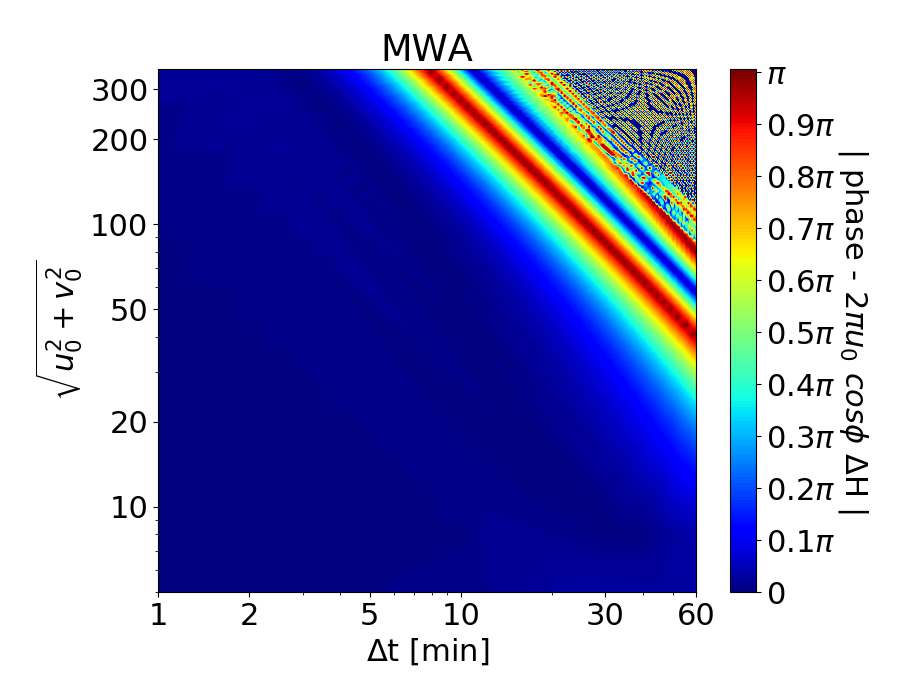}
	\end{minipage}\hfill
\centering
	\begin{minipage}{0.49\textwidth}
		\centering
		\includegraphics[width=1.0\linewidth]{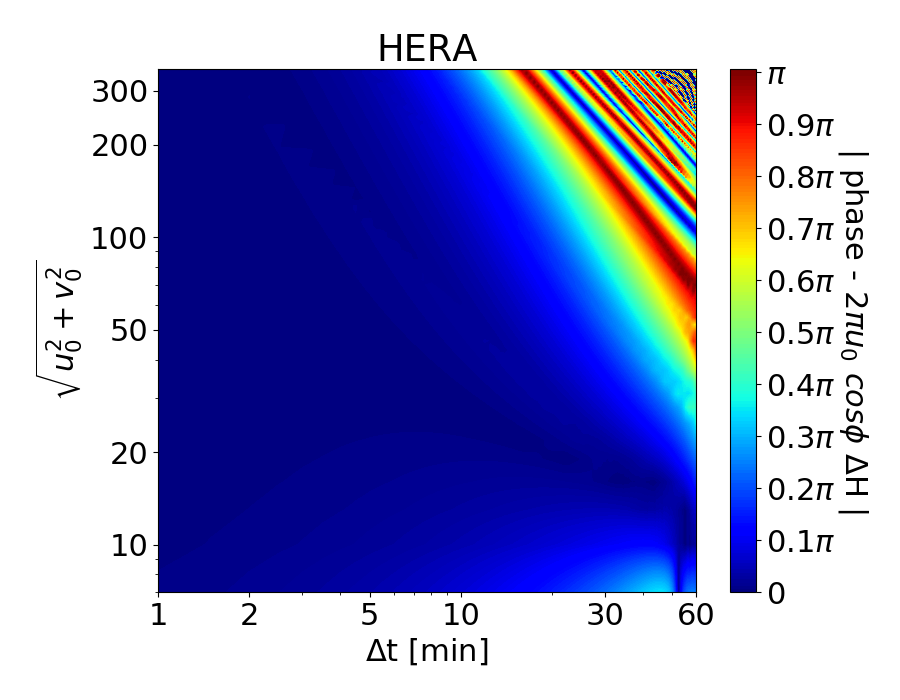}
	\end{minipage}\hfill
	\centering
	\begin{minipage}{0.49\textwidth}
		\centering
		\includegraphics[width=1.0\linewidth]{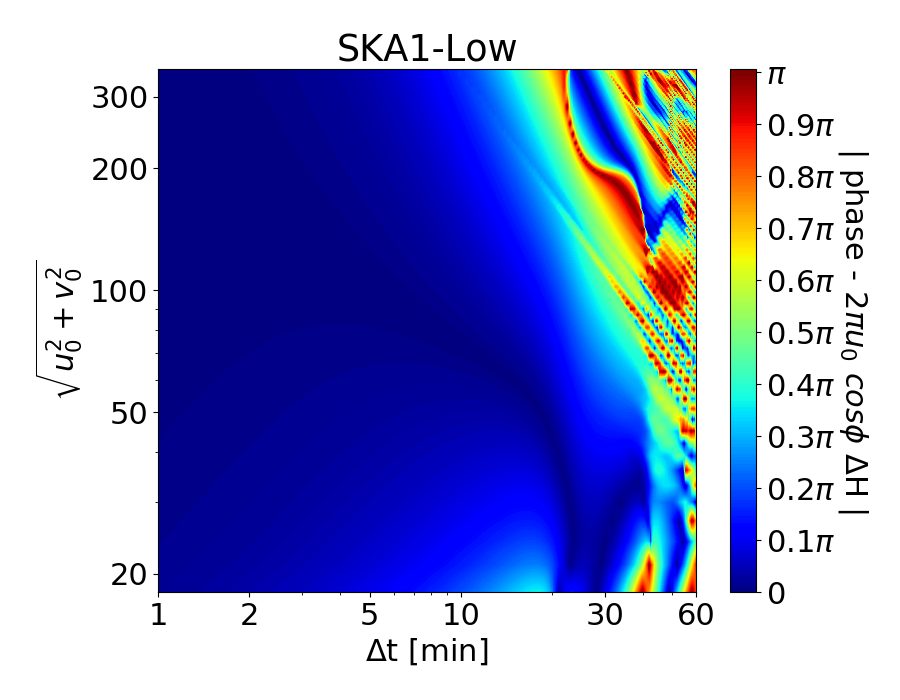}
	\end{minipage}\hfill
\caption{ 
The figure shows the absolute value of the phase angle of the visibility correlation function (Eq.~(\ref{eq:cov_phase})) as a function of $\Delta t = t'-t$.  This figure illustrates that the rapidly fluctuating component of the phase of the complex correlation function (Eq.~(\ref{eq:viscorapprox5})) can mostly be removed  by multiplying  it   with $\exp(-i2\pi u_0 \cos{\phi} \Delta H)$. This allows us to determine the   time scales for averaging  the time-ordered visibilities  in  drift scans (section~\ref{sec:phase} and~\ref{sec:ana_data_dr}). 
 }
\label{fig:HIcorr_cov_ph}
\end{figure}

\subsubsection{The phase of visibility correlation function} \label{sec:phase}
In the foregoing we  studied  the amplitude of 
the correlation function. As the correlation function (in either frequency or 
delay space Eq.~(\ref{eq:viscorapprox2}) or Eq.~(\ref{eq:visdelaysp})) is a  complex function we need to know the correlation properties of its phase in addition to complete the analysis. 

In Appendix~B, we discuss how  suitable approximations allow us to discern major contributors  to the phase of the correlation function.  Eqs.~(\ref{eq:viscorr_delta1}) and~(\ref{eq:phase_gua}) show that the phase angle is $2 \pi u_{0} \cos{\phi} \Delta H + \psi_1 + \psi_2 $. The term $2 \pi u_{0} \cos{\phi} \Delta H$ has already been discussed above (point~(a) on traversal time of coherence scale).  It follows from Eq.~(\ref{eq:phase_gua}) that both $\psi_1$ and $\psi_2$   are  small  as compared to $2 \pi u_{0} \cos{\phi} \Delta H$ as  $\psi_1 \propto \Omega_g$ and $\psi_2 \propto \Omega_g^2$  for $\pi^2 \Omega_g^2 y^2 < 1$.  $\psi_2$  can only be significant    when effects arising from large field-of-view  become important (Eq.~(\ref{eq:phase_gua}) and discussion on point~(c) above), which is not the case  for $w_0 =0$ and  the  primary beams we consider in our analysis. The dominant  phase angle  $2 \pi u_{0} \cos{\phi} \Delta H$  can be  explicitly identified  in Eq.~(\ref{eq:viscorr_delta1}) in this case.

Motivated by our analytic results, we define the phase angle  as:
\begin{equation}
\psi(\textbf{u}, t'-t) = {\rm Arg}\left(\exp(-i2\pi u_0 \cos{\phi} \Delta H) \Big\langle V_{\tau}(\textbf{u}_{0}, w_{0},t)V^{*}_{\tau}(\textbf{u}_{0}, w_{0},t') \Big\rangle \right )
\label{eq:cov_phase}
\end{equation}
The multiplication by the  additional phase  allows for near cancellation of the phase term  $\exp(ik_{\perp 1} r_0 \cos{\phi} \Delta H)$ in Eq.~(\ref{eq:visdelaysp}) (or a similar term in Eq.~(\ref{eq:viscorapprox2}) for 
correlation in frequency space if $u_0$ and $r_0$ are  replaced by $u_\nu$ and 
$r_\nu$, respectively). In Figure~\ref{fig:HIcorr_cov_ph} we present our 
numerical results.    We notice that the phase angle defined by Eq.~(\ref{eq:cov_phase}) is small for a wide range of $\Delta t$, as suggested by our analytic results. 
This means, as anticipated, that  the phase  of the correlation function is nearly $\exp(i2\pi u_0 \cos{\phi} \Delta H)$ \footnote{The origin of this phase can  partly be 
explained 
by considering a  simpler case: a single point source of flux $F_\nu$  at the phase center. In this case, the visibility $V_\nu(\textbf{u}) = F_\nu A_\nu(0)$, where $A_\nu(0)$ defines the primary beam response at the phase center, $l=0$ and  $m=0$.  The  correlation between visibilities separated by $\Delta H$  in  time in 
a drift scan is $V_\nu(\textbf{u}) V^*_\nu(\textbf{u}) \simeq  F_\nu^2 A_\nu^2(0) \exp(i2\pi u_\nu \cos{\phi} \Delta H)$. As discussed in section~\ref{sec:psources}  the same factor  scales out of the correlation function 
for a set of point sources also.}. The implication of this result for drift scan data analysis  will be discussed below.

\subsection{Approximations and input quantities} \label{sec:app_inp}
Our results use an input HI power spectrum, different primary beams, and 
a set of approximations to transform from frequency to delay space. We discuss
the impact of these approximations and input physics on our analysis.

\subsubsection{Dependence on input power spectrum and the shape of primary beam} \label{sec:pk_effect}
The results shown in Figure~\ref{fig:HIcorr2} were derived using the HI power 
spectrum, $P(k) \simeq 1/k^n$, with $n\simeq 2$, for a range of scales (\citealt{furlanetto06}). 
We tested our results with different power-law  HI  power spectra with  spectral indices in the range  $n = 1\hbox{--}3$
and found  our results to be insensitive  to the input power spectra. 

The  lack of dependence of the visibility decorrelation time  on the input HI  power spectrum follows from our analysis. Eqs.~(\ref{eq:viscorr_delta}) and~(\ref{eq:viscorr_delta1}) show that relevant approximations allow us to separate the input power spectrum from the time-dependent part of the correlation function, which means Figure~\ref{fig:HIcorr2} is independent of the HI power spectrum. These equations show that the time dependence   of the correlation function is  
essentially captured by the response of the primary beam in Fourier space. 
Similar expression was derived in \cite{fringe-rate} (their equation~9) for cases when  the Fourier  beam (Eq.~(\ref{eq:viscorapprox3})) has a narrow response (e.g. PAPER).

 The only cases not covered by this approximation are small primary beams
and small baselines.   However, for the limiting cases  we discuss here, $|\textbf{u}| \ga 20$ and SKA1-Low primary beam, our numerical results show that the impact of the input HI power spectrum on the decorrelation time scale is negligible.

Our results are insensitive to the shape of the primary beam. We compare our numerical 
results for  instrumental primary beams
with  a symmetric,   separable Gaussian beam by  roughly matching     $\Omega_{0g}$ and the  main lobe of the instrumental primary beam. We find excellent agreement 
in explaining the main features of Figures~\ref{fig:HIcorr2}, \ref{fig:HIcorr_cov} and~\ref{fig:HIcorr_cov_ph}. Eq.~(\ref{eq:viscorr_delta1}) adequately explains Figure~\ref{fig:HIcorr2}, except for small baselines
for HERA and SKA1-Low.

\subsubsection{Approximations in transforming from frequency to delay space}
Following Eq.~(\ref{eq:visdelaysp}) we discuss various approximations used in making  
the correlation function in delay space more tractable. In the tracking case,
these  approximations allow us to find a one-to-one linear  relation between  the Fourier modes of the HI signal
with the variables of radio interferometers (e.g. \citealt{paul16} and references therein).   However, owing to  the frequency dependence of the primary beam, the coordinate distance, and the baseline, these commonly-used  relations  are approximate. We assessed  the impact of these approximations in \cite{paul16} for the  tracking case.  For a bandwidth $B = 10$~MHz ($\nu_{0}=154$~MHz) and  MWA primary beam, the error in these relations 
  is less than 5\% for $k_\parallel \ga 0.1 \, \rm Mpc^{-1}$. The modes corresponding to $k_\parallel \la 0.1 \, \rm Mpc^{-1}$  are buried in the foreground wedge and therefore do not 
play a role in the detection of the HI signal (e.g. \citealt{paul16}).  The error increases with
bandwidth and primary beam and therefore  is expected to be smaller for HERA and SKA1-Low for the same bandwidth. As we also use these approximations in our work  to separate the variables on the sky-plane from  those along the line-of-sight, we re-assess these approximations for a drift scan and  find these errors to be of similar 
magnitude for the drift scan. As in the tracking case, these approximations allow us to derive the relation between baseline and delay space parameter $\tau$ and Fourier modes of the HI signal.  This simplification  allows us
to write the frequency-dependent terms in the form expressed  in Eq.~(\ref{eq:viscorapprox4}). 

One outcome of this approximation for drift scans is  that the functional form of the decorrelation time  shown in  Figure~\ref{fig:HIcorr2} is nearly the same in frequency and delay space. Therefore, Figure~\ref{fig:HIcorr2} can be interpreted as displaying the 
decorrelation time  at the center of the bandpass. 
This assertion is borne out by Eq.~(\ref{eq:viscorr_delta}).

 Our study is based on the assumption    $\nu_0 \simeq 154 \, \rm MHz$ and $B \simeq 10 \, \rm MHz$. It   can  readily be extended
to a different  frequency/bandpass by using  
Eqs.~(\ref{eq:viscorr_delta}) and/or~(\ref{eq:viscorr_delta1}). 

We discuss the approximation in transforming from frequency to delay space 
further with regard to foregrounds and  the analysis of drift scan data in  later sections (see footnote~\ref{fn:delays}). 

It is worthwhile to reiterate the scope of the main  approximations we use:
(a) For large primary beams and baselines, Eq.~(\ref{eq:viscorr_delta}) provides
an excellent approximation, (b) for  small bandwidths and  primary beams, 
Eq.~(\ref{eq:viscorr_delta}) can readily be extended to Eq.~(\ref{eq:viscorr_delta1}), (c) for small baselines and primary beams, Eq.~(\ref{eq:viscorr_delta})
might not be  valid and Eq.~(\ref{eq:viscorapprox5}) has to be computed numerically.

\section{Foregrounds in drift scans} \label{sec:fore_drift} 

In the tracking mode, the foregrounds can be isolated from the HI signal (`EoR window') by transforming to delay space if the two-dimensional foregrounds are spectrally smooth and therefore their correlation scales differ  from  the   three-dimensional  HI signal along the line of sight. However, in tracking mode, we 
cannot use the difference between correlation properties of foregrounds and the  HI 
signal on the sky plane. In a drift scan, it is possible that the  
decorrelation time of the HI signal is different from components of foregrounds, which  might  give  us yet another way to mitigate foregrounds. 

The aim of this section is to study the decorrelation time scales of 
two components of foregrounds: near-isotropic distribution of point sources 
of flux above $1$~Jy and  statistically homogeneous and isotropic diffuse 
foregrounds. In our analysis,  the delay space approach   continues to be the primary method used to isolate foregrounds from the HI signal and we therefore  present all our results in this space.

\subsection{Point Sources} \label{sec:psources}
In a drift scan the phase center is held fixed while the intensity pattern changes. 
The changing intensity pattern owing to a set of point sources can be written as: 
\begin{equation}
I_\nu({\boldsymbol \theta},t) = \sum_m F_\nu^m \delta^2({\boldsymbol \theta} - {\boldsymbol \theta_m}(t))
\end{equation}
 Here  $F_\nu^m$ is the flux of the $m^{\rm th}$  source and   ${\boldsymbol \theta_m}(t)$ its  angular position  at time $t$.
Here all the angles are measured with respect to the phase center  which 
is assumed to be fixed at  ${\boldsymbol \theta_0} = 0$. The visibility (retaining the $w$-term) can readily be derived from
the expression above:
\begin{equation}
V_\nu({\bf u}_\nu, w_{\nu}, t) = \sum_m F_\nu^m A_\nu({\boldsymbol \theta}_m(t)) \exp \left[ {-2{\pi}i\left({\textbf{u}_{\nu}}{\cdot}{\boldsymbol \theta}_m(t) + w_{\nu}(n_m(t)-1) \right)} \right] \label{eq:vis_pt_f}
\end{equation}
To discern  the main results of  this section, we  ignore the frequency dependence of source fluxes  and primary beam, even though we allows these  quantities to be frequency dependent in our simulations
\footnote{We  neglect the frequency dependence of the intensity pattern and the primary beam throughout this 
paper. As we compare our analytic results against simulations in this section, it allows us to verify this assumption more explicitly. We find this assumption to be extremely good for bandwidth $B\simeq 10 \, \rm MHz$ around a central frequency of $\nu_0\simeq 154 \, \rm MHz$. This approximation can be understood by
considering a simpler case: a flat spectrum source at the phase center. While transforming to delay space,  this source receives contribution from only the   $\tau=0$ mode. If the source is now assumed to have a spectral index, more delay space modes close to $\tau=0$ begin to contribute. We find that  these modes do not contaminate the EoR window as they 
lie well within the wedge given  the bandwidth and spectral index of interest. The leakage into the EoR window owing to finite bandwidth can be assuaged by 
using a frequency-space convolving function  such as Blackman-Nuttall window or a Gaussian
window we discuss in the section on diffuse foregrounds.   The  frequency dependence  of baselines  in the phase   plays a more important role and is needed to explain the wedge structure for foregrounds (e.g. \citealt{paul16}).\label{fn:delays}}.
Using Eq.~(\ref{eq:defdelay}) the visibility of point sources in the delay space is: 
\begin{align}\label{eq:vis_pt_tau}
V_\tau({\bf u}_0,w_0,t) &\simeq \sum_m F_0^m A_0({\boldsymbol \theta}_m(t)) B \sinc(\pi B \bar{\tau}^m (t)) e^{2 \pi i \nu_0 \bar{\tau}^m (t) } \\
\text{where,} \qquad 
\bar{\tau}^m (t) &= \tau - \frac{1}{\nu_0} \left( {\textbf{u}_0}{\cdot}{\boldsymbol \theta}_m(t) + w_0 (n_m(t)-1) \right)
\end{align}
The correlation function of the visibilities in delay space can be written as:
\begin{align}
	\big\langle V_\tau({\bf u}_0,w_0,t) & V^*_\tau({\bf u}_0,w_0,t') \big\rangle \simeq B^2 \sum_m \sum_n F_0^m F_0^n A_0({\boldsymbol \theta}_m(t)) A_0({\boldsymbol \theta}_n(t')) \nonumber \\
	&\times \sinc(\pi B \bar{\tau}^m (t)) \sinc(\pi B \bar{\tau}^n (t')) e^{2 \pi i \nu_0 \left( \bar{\tau}^m (t) - \bar{\tau}^n (t') \right) } \label{eq:forepow}
\end{align}
Here the ensemble average implies averages over all pairs of baselines and times for  which $|{\bf u}_0|$ and $t'-t$  are held fixed.
To understand  Eq.~(\ref{eq:forepow})  we first consider the tracking case in which source positions are independent of time. In this case the dominant contribution comes from $\tau = 2\pi {\bf u}_0.{\boldsymbol \theta_m}/\nu_0$. This defines the so-called foreground wedge which is bounded by the maximum value of $\theta_m$ which is given approximately by the size of the  primary beam.  It also follows from the equation  that the sum is dominated by terms for which $m=n$. 

In a drift scan the source position changes with respect to the primary beam. It means the value of $\tau$ for which the sum in Eq.~(\ref{eq:forepow})  peaks changes with time. While the broad wedge structure is the same in this case as in the tracking case as the dominant contribution comes from sources within the primary beam, the correlation structure  becomes more complicated.   As ${\boldsymbol \theta_n}(t') - {\boldsymbol \theta_m}(t)$ remains unchanged  during a drift scan, the summation in this case would  also generally be   dominated by $m=n$ terms. However, it is possible that a source at one position at a time drifts close to the position of another source at another time. Even though the contribution of this pair could be negligible in tracking mode, it would not be if the visibilities are correlated at two different times. The impact of this effect requires details of point source distribution which we  model using a simulation in this paper. 

For the case of $m=n$, the same source is correlated at two different 
times. In this case,  it follows from Eq.~(\ref{eq:forepow}) that the visibility correlation diminishes as the time separation increases. As the additional time-dependent  phase acquired in the drift is  proportional to the length of the baseline, the decorrelation time scale is expected to be shorter  for longer  baselines.

\textbf {\textit{Point source simulations}}: 
We generate  15067 point sources brighter than $1\, \rm Jy$ distributed isotropically on the southern hemisphere (\citealt{hopkins03}). We assume the 
spectral index of sources to be $-0.7$\footnote{Foreground components from both the point sources and diffuse galactic emission are expected to be dominated by synchrotron radiation from power-law energy distribution of relativistic electrons. The galaxy is optically thin to these photons, therefore, the observed spectrum retains the form of the emitted spectrum, which is featureless. The main mechanism  of the absorption of radio photons
in the interstellar medium is free-free absorption off  thermal and non-thermal electrons. The optical depth of free-free absorption: $\tau = 3.3\times 10^{-7} (T/10^4)^{-1.35} \nu^{-2.1} \rm EM$, where $\nu$ is in GHz and EM, the emission measure, is observationally determined to be: $EM = 5 \, \rm pc \, cm^{-3}$ (e.g. \citealt{1999ApJ...523..223H});  the optical depth  is negligible at frequencies of interest to us.} For this source distribution we compute
the power spectrum in delay space as a function of drift time. 
In a  drift scan,  the  coordinates of these sources evolve according to Eq.~(\ref{eq:dldmdn}) with respect 
to the fixed phase center.

We  compute visibilities in delay space for  a one-hour drift scan.  The visibilities are then correlated in time and  the  visibility correlation function is computed by averaging over  the    number of correlation pairs for which $t'-t$  and $|{\bf u}_0|$ are   held fixed: 
\begin{align}\label{eq:def_corr_fn_pt}
	 \big\langle V_\tau({\textbf u}_0,w_0,t) & V^*_\tau({\textbf u}'_0,w'_0,t') \big\rangle = 
	 \frac{1}{N_{|{\bf u}_0|}} \sum_{|{\bf u}_0|} ^{N_{|{\bf u}_0|}} \frac{1}{N_{tt'}} \sum_{t, t'} ^{N_{tt'}} V_\tau({\bf u}_0,w_0,t) V^*_\tau({\bf u}'_0,w'_0,t')
\end{align}
Here $N_{|{\bf u}_0|}$ and $N_{tt'}$ and the number of baseline pairs for fixed $|{\bf u}_0|$ and $t-t'$, respectively.

To establish how  the amplitude of the  visibility correlation  behaves as a function of time,  baselines,  and 
the number of points over which the average is computed, we choose two representative baselines  $|{\bf u}_0| = 20, 100$. We carry out averages in a ring of width $\Delta |{\bf u}_0| = 4$; each of these rings is populated, randomly and uniformly, with   $N_{|{\bf u}_0|}=  25, 50, 100, 200, 400$. 

\begin{figure}[ht]
	\centering
	\begin{minipage}{0.49\textwidth}
		\centering
		\includegraphics[width=0.99\linewidth]{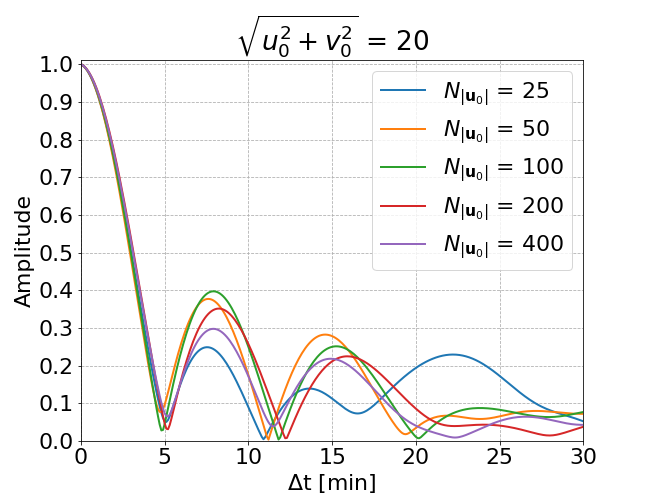}
	\end{minipage}\hfill
	\centering
	\begin{minipage}{0.49\textwidth}
		\centering
		\includegraphics[width=0.99\linewidth]{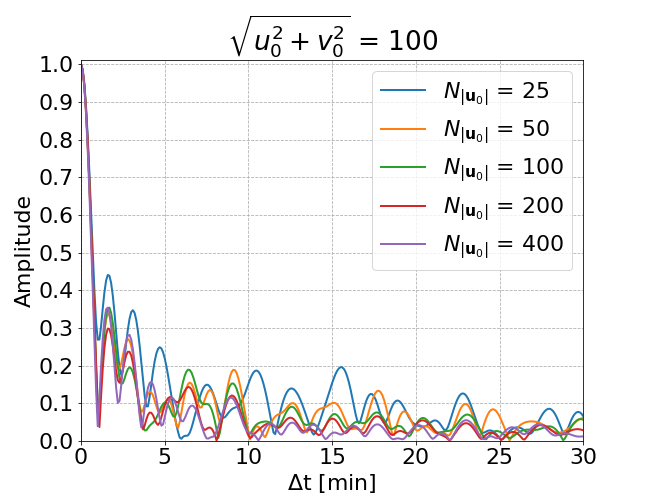}
	\end{minipage}\hfill
	\caption{ 
	The visibility correlation function  (Eq.~(\ref{eq:def_corr_fn_pt})) is shown  as a function of $\Delta t$ (normalized to unity for $\Delta t =0$) for two baselines $\sqrt{u_0^2+v_0^2} = 20, 100$ for $u_0=v_0$, for MWA primary beam and latitude.  The visibility correlation function is seen to fall to half its value  in   a few  minutes. 
}
	\label{fig:pt_corr0203_1}
\end{figure}
In Figure~\ref{fig:pt_corr0203_1}, the visibility correlation functions are plotted for the two cases using the instrumental parameters of MWA (primary beam and $\phi$) for $\tau=0$ and $w_0=0$.   We notice the following: (a) averaging over more baselines causes the correlation function to decorrelate faster  when the number of baselines are small but the function converges as the number of baselines is increased, (b) the correlation function 
decorrelates faster for larger baselines, as anticipated earlier in the section  based on the analytic expression, Eq.~(\ref{eq:forepow}), (c) a comparison between  
Figures~\ref{fig:pt_corr0203_1} and~\ref{fig:HIcorr2} shows the decorrelation time scale for the HI signal is much larger  than for a set of point sources. 
For $|{\bf u_0}| = 100$, the point sources decorrelate to 50\% of the peak in less than a minute while this time is nearly 10~minutes for the HI signal. 

The structure of the foreground wedge in a drift scan is expected to be 
similar  to the  tracking mode; we verify it using analytic estimates and simulations but do not show it here.

\subsection{Diffuse correlated foregrounds}

An important contribution to the foregrounds comes from diffuse galactic emission (DGE) which is correlated on the sky plane; this component of the  foregrounds is dominated by optically-thin galactic synchrotron emission. The spatial and 
frequency dependence of this emission is separable if the emission is optically thin, which, as noted above,  is  a good assumption   and is key to the separation of 
foregrounds from the HI signal.  We consider  statistically homogeneous and isotropic  component of the diffuse foreground here. This case differs from  
 the HI signal only in  different frequency dependencies of the two signals. 
Therefore, the  formulation is similar  to the case of HI signal discussed above.

As we assume the DGE  to be statistically homogeneous and isotropic, the two-point function of  fluctuations  on the plane of the sky  in Fourier space could be  characterized by  a power spectrum $C_q$ such that   and $q = |{\textbf q}| = \sqrt{q_1^2 + q_2^2}$, where ${\textbf q} = (q_1, q_2)$, with $q_1$ and $q_2$ being   the two
Fourier components on the sky plane. $C_q$ can be expressed  as:
\begin{equation}
  \big \langle I_\nu( {\textbf q} ) I_{\nu'}( {\textbf q}' ) \big \rangle 
  = (2 \pi)^2 C_q (\nu, \nu') \delta^2 ( {\textbf q} - {\textbf q}' ) 
\end{equation}
For our analysis we adopt the following form and normalization of  $C_q$, as appropriate for $\nu \simeq 150$ (e.g. \cite{ghosh12} and references therein):
\begin{equation}
  C_q (\nu, \nu') = a_0 \left (\frac{\nu}{\nu_0} \right)^{-\alpha } \left(\frac{\nu'}{\nu_0} \right)^{-\alpha } \left(\frac{q}{1000} \right)^{-\gamma } \label{eq:DGSE_ps}
\end{equation}
 where $\alpha = 0.52$ (\citealt{rogers08}) is the spectral index and $\gamma = 2.34$ (\citealt{ghosh12}) is  the index of spatial power spectrum. The value of $a_0 = A_0 \left( 2 k_B \nu_0^2 / c^2 \right)^2$ is 237~${\rm Jy}^2$ at $\nu_0 = 154 \, {\rm MHz }$. It rescales the amplitude factor, $A_0 = 513 {\rm mK}^2 $,  given in \cite{ghosh12} from $({\rm mK})^2$ at 150~MHz to ${\rm Jy}^2$ at $\nu_0$. For a single polarization this factor should be divided by 4. 

Using the formalism used for analysing the HI signal it can readily be shown that the visibility correlation function in frequency space can be related to $C_q$ as:
\begin{align}
 \Big\langle V_{\nu}(\textbf{u}_{\nu}, w_{\nu},t)V^{*}_{\nu'}(\textbf{u}'_{\nu'}, w'_{\nu'},t') \Big\rangle 
 &= \int \frac{d^2 q}{ (2 \pi)^2} C_q (\nu, \nu') e^{i  q_1 \cos{\phi} \Delta H} 
 Q_{\nu}(\textbf{q}, \textbf{u}_\nu, w_{\nu}, \Delta H = 0) 
 Q^*_{\nu'}(\textbf{q}, \textbf{u}'_{\nu'}, w'_{\nu'}, \Delta H)
\label{eq:diff_corrfun}
 \end{align}
where the Fourier beam of DGE is:
 \begin{align}
 Q_\nu(\textbf{q}, \textbf{u}_{\nu}, w_{\nu}, \Delta H) &= {\int} d^{2}\theta A_{\nu}(\boldsymbol{\theta}) 
	\exp\left[ -2 \pi i \left( \boldsymbol{x}_u \cdot {\boldsymbol{\theta}} -\frac{1}{2} y \theta^{2}\right) \right] \label{eq:DGSE_Q} \\
\text{with} \mkern+200mu
	x_u &= u_{\nu} - \frac{1}{2\pi} \left(q_1 + q_2 \sin{\phi} \Delta H \right) \\
	x_v &= v_{\nu} - \frac{1}{2\pi} \left(q_2 - q_1 \sin{\phi} \Delta H \right) \\
	y   &= w_{\nu} + \frac{1}{2\pi} q_1 \cos{\phi} \Delta H
\end{align}
In  Eq.~(\ref{eq:DGSE_Q}) we have used  $Q$-integrals (or 2D Fourier beam) defined for the HI  correlation function (Eq.~(\ref{eq:viscorapprox3})). Comparing Eq.~(\ref{eq:DGSE_Q}) and 
Eq.~(\ref{eq:viscorapprox3}) we note that the following relation between the  Fourier modes of correlated diffuse foregrounds and the HI signal: $ {\bf q} \simeq r_0 {\bf k_\perp}$. 

As already shown for the HI signal, Eq.~(\ref{eq:DGSE_Q})  can be made  more tractable by assuming the primary beam to be  separable and symmetric.  To establish  general characteristics of DGE foreground we carry out  analytical calculations with a symmetric Gaussian beam: $e^ {- {(l^2 + m^2)}/{\Omega_g}}$, which allows us to extend the integration limits from $-\infty$ to $+\infty$.  
Following the HI analysis, we also expand $n$ to the first order. This gives us:
\begin{align}
 Q_\nu(\textbf{q}, \textbf{u}_{\nu}, w_{\nu}, \Delta H) &= \pi \Omega'_g \exp \left[ - \pi^2 \Omega'_g \left( x_u^2 + x_v^2 \right) \right] \label{eq:Fourier_beam_freq}
\end{align}
where $ \Omega'_g = \Omega_g / (1 - i\pi y \Omega_g)$. 
It should be noted that these variables can be read off directly from $Q$-integrals defined for the HI signal by putting $r_0 {\bf k_\perp} \simeq {\bf q} $. This
shows the equivalence of the HI signal and diffuse foregrounds in 
the Fourier domain on the plane of the sky.  

We next carry out frequency integrals to transform to delay space. As already discussed in section~\ref{sec:psources}, the main results in  the delay space can be obtained  by retaining only  the frequency dependence of   baselines because   the foregrounds  wedge in the two-dimensional  power spectrum of foregrounds arises  largely due to  the chromaticity of  baselines (e.g. \citealt{paul16}). 

The frequency integral can be computed  numerically for a  finite bandpass. 
To carry out analytical calculations,  the limits of the frequency integral can be extended to infinity.  However,  under this assumption, the baseline ($ {\bf u}_\nu = {\bf u}_0 \nu / \nu_0$) also becomes infinity and the integral does not converge 
\footnote{This highlights the main difference between the HI signal  and the  two-dimensional diffuse foregrounds. In the former, the frequency
integral picks the scale along the line-of-sight $k_\parallel$ while no such
scale exists for diffuse foregrounds}. 
To correctly pick the relevant scales
 of diffuse foregrounds,  we  apply a Gaussian window function in frequency space ($\exp \left(-c_2 (\nu-\nu_0)^2 \right)$) which  allows us to pick the relevant scales within the  bandwidth ($B$) of the instrument and also enables us to extend the limits of integration. \footnote{A similar window (e.g. Blackman-Nuttall window, e.g. \citealt{paul16}) is applied  to the data to prevent the leakage  of foregrounds
from the foreground wedge to the clean EoR window.}
 This gives us:
\begin{align}
	\tilde Q(\textbf{q}, \textbf{u}_0, w_0, \Delta H) = 
	\int_{\nu_0 - B/2}^{\nu_0 + B/2} d\nu  e^{2 \pi i \tau \nu} e^{ -c_2 (\nu-\nu_0)^2 } Q_\nu(\textbf{q}, \textbf{u}_{\nu}, w_{\nu}, \Delta H)	\nonumber \\
	= \pi \Omega'_g \sqrt{\frac{\pi}{c_1 + c_2}} \exp \left[-\frac{\pi^2 \tau^2}{c_1 + c_2} \right] \nonumber 
	\exp \left[ 2\pi i \tau \nu_0 \left(1 + \frac{c_1}{c_1 + c_2} \frac{1}{ |{\bf q_u}| } \left( a_1 + a_2 \sin{\phi} \Delta H  \right) \right)  \right] \nonumber \\
	\times \exp \left[ -\frac{\Omega'_g}{4} \left( \frac{c_2}{c_1 + c_2} \left( a_1 + a_2 \sin{\phi} \Delta H  \right)^2 + 
	\left( a_2 - a_1 \sin{\phi} \Delta H - |{\bf q_u}| \sin{\phi}\Delta H \right)^2
	\right) \right] \label{eq:Fourier_beam_delay}
\end{align}
where $c_1 = \left( |{\bf q_u}| / \nu_0 \right)^2 \Omega'_g / 4, c_2 = 1/(b B^2), {\bf q_u} = 2\pi {\bf u}_0, a_1 = q_1 - 2\pi u_0, a_2 = q_2 - 2\pi v_0 $.  The parameter  $b$ is a numerical factor which can be tuned to get the desired width of the Gaussian window function.  The argument of the  factor $\exp \left[- {2 \pi^2 \tau^2} / ({c_1 + c_2}) \right]$ in Eq.~(\ref{eq:Fourier_beam_delay}) yields the linear relation corresponding to the foreground wedge. 
\begin{figure}
	\includegraphics[width=0.49\linewidth]{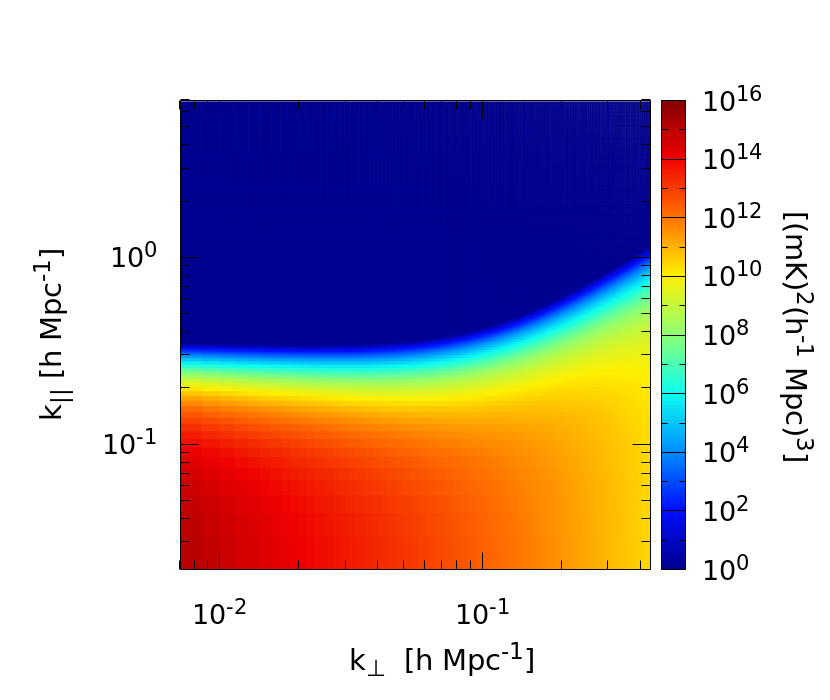}
	\includegraphics[width=0.49\linewidth]{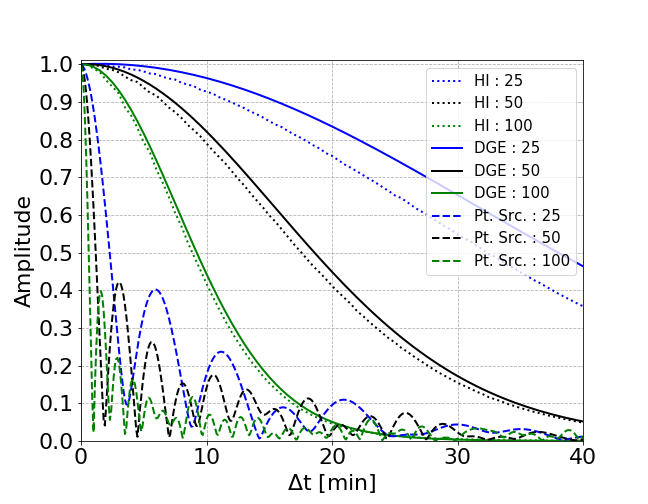}
	\caption{ In the left panel,  we show two-dimensional  power spectrum of DGE ($\Delta H =0$) in the   $k_\parallel\hbox{--}k_\perp$ plane in units 
$\rm (mK)^2 (h^{-1} Mpc)^3$. The figure assumes $\nu_0 = 154 \, \rm MHz$ and bandwidth $B=10 \, \rm MHz$. The    
relation applicable to the HI signal  is used to transform from  the telescope variables ($u_0,v_0$,  $\tau$) to the Fourier modes  
(${\bf k}_\perp$,  $k_\parallel$),   and to convert  the power spectrum to the appropriate units (e.g. \citealt{paul16}). The Figure highlights  the separation of foregrounds from the EoR window;  the bandwidth  determines the extent of the flat region parallel to the  $k_\parallel$ axis.  
In the right panel,  the  visibility correlation function (normalized to unity for  $\Delta t = 0$)  for  DGE is shown  for three baselines
$\sqrt{u_0^2 + v_0^2} = 25, 50, 100$ (Eq.~(\ref{eq:DGSE_viscorr})). We also show the HI and point source visibility correlation functions for comparison. 
	}
	\label{fig:DGSE}
\end{figure}

We can read off the correlation scales for diffuse correlation foregrounds
from Eq.~(\ref{eq:Fourier_beam_delay}). A baseline ${\bf u}_0$ is  most sensitive to the Fourier mode ${\bf q_u}$.   As in the case of the HI signal, the decorrelation  time scale for a drift scan  can be estimated readily by putting ${\bf q} = {\bf q}_u$ and simplifying the expression. We finally obtain: 
\begin{align}
	\Big\langle V_\tau({\bf u}_0,w_0,t) & V^*_\tau({\bf u'}_0,w'_0,t') \Big\rangle = \iint_{\nu_0 - {B} / {2}}^{\nu_0 + {B} / {2}} d\nu d\nu'
	\Big\langle V_{\nu}(\textbf{u}_{\nu}, w_{\nu},t)V^{*}_{\nu'}(\textbf{u}'_{\nu'}, w'_{\nu'},t') \Big\rangle \nonumber \\
	&= \int \frac{d^2 q}{ (2 \pi)^2} C_q (\nu_0, \nu_0) e^{i  q_1 \cos{\phi} \Delta H}  \tilde Q (\textbf{q}, {\bf u}_0, \Delta H = 0) \tilde Q^* (\textbf{q}, {\bf u}'_0, \Delta H) \label{eq:DGSE_viscorr}
\end{align}
Eq.~(\ref{eq:DGSE_viscorr})   gives the general expression for  visibility correlation function in delay space for a drift scan observation. It can be computed by using Eqs.~(\ref{eq:DGSE_ps}), (\ref{eq:Fourier_beam_delay}) in Eq.~(\ref{eq:DGSE_viscorr}).  It reduces to the  relevant expression   for  tracking observation for  $\Delta H = 0$. 
In Figure~\ref{fig:DGSE} we show numerical  results obtained from solving Eq.~(\ref{eq:DGSE_viscorr}) for a Gaussian primary beam matched to the main lobe of MWA primary beam and $\phi = -26.7^\circ$.  We display the power spectrum  in $k_\parallel\hbox{--}k_\perp$ plane for $\Delta H = 0$  and
the correlation  of  diffuse correlated foregrounds as a function of time. Our main conclusions are:
\begin{itemize}
\item[1.] Like the point sources, diffuse correlated foregrounds are confined to a wedge and the EoR
window is clean for the detection  of the HI signal. 
\item[2.]  The  diffuse  foregrounds decorrelate on time scales comparable to  the HI signal. (We note that the  difference between the two cases
  for the shortest baseline is partly because we use the exact MWA beam for the HI case while we use the Gaussian beam for diffuse foreground.)   This should be contrasted with  point-source foregrounds that decorrelate on a much shorter time scale as compared to the HI signal. 

\end{itemize}

\section{Analysing drift scan data}  \label{sec:ana_data_dr}

Our study  allows us to address the following question: over what time period can the time-ordered  visibility data be  averaged without diminishing  the  HI signal. 
We further seek optimal signal-to-noise for the detection of the HI signal. 
We computed two-point visibility correlation function     to assess  the coherence time scale 
of visibilities.  Our results are shown  in Figures~\ref{fig:HIcorr2}--\ref{fig:HIcorr_cov} (amplitude of the correlation function as a function of $\Delta t$,  baseline and primary beam)  and~\ref{fig:HIcorr_cov_ph} (the phase of the complex  correlation function).  Our study shows that  the range of  time
  scales over which time-ordered visibilities can be averaged without the loss
  of HI signal lies in the range of a few minutes to around 20 minutes. 

Motivated by our theoretical analysis, we  define the quantity:
\begin{equation}
{\cal C}_\tau(\textbf{u}_{0},w_0,t'-t) \equiv \exp(-i2\pi u_0 \cos{\phi} \Delta H) \Big\langle V_{\tau}(\textbf{u}_{0}, w_{0},t)V^{*}_{\tau}(\textbf{u}_{0}, w_{0},t') \Big\rangle
\label{corr_def_phase}
\end{equation}
Notice that ${\cal C}_\tau(\textbf{u}_{0},w_0,t'-t) = {\cal C}^*_\tau(\textbf{u}_{0},w_0,t-t')$. Our analysis shows that the complex number  ${\cal C}_\tau(\textbf{u}_{0},w_0,t'-t)$ is dominated by its  real component with a   phase which remains small over  the 
coherence time scale  of  the amplitude  (Figure~\ref{fig:HIcorr_cov_ph} and  Figure~\ref{fig:HIcorr_cov}). Our aim is to extract ${\cal C}_\tau(\textbf{u}_{0},w_0,t'-t)$ from the data and then suitably weigh it to extract the 
HI signal, optimally and without the loss of HI signal\footnote{To prevent HI signal loss, the simplest way to extract the HI signal from drift scans would be to not use the coherence
of visibilities in time.  Assuming visibilities are measured with time resolution much shorter than the coherence time scale,  visibilities with identical time stamps 
can be squared (after averaging over redundant baselines) to compute the power spectrum. This gives an unbiased estimator of the HI signal. However, in such a procedure, visibilities  measured at two different times
are treated as uncorrelated which results in an   estimator with  higher noise as compared to what is achievable using further information regarding coherence of visibilities in time. If the time resolution of  visibilities is around 10~seconds and the coherence time is around 10~minutes, then the  noise RMS of the  
visibility correlation  is higher by roughly the square root of the  ratio of these two times.}. 
We discuss two possible ways to extract the HI signal. The first is based on 
averaging the visibilities before computing the correlation function.

We consider visibilities measured with  time resolution  $\Delta H$ ($\Delta H$ is assumed to be much smaller than the coherence scale of visibilities for any baseline of interest to us, e.g. $\Delta H =10\, \rm sec$). Let us denote the measured  visibilities as,  $V_n$, where $n$ corresponds to the time stamp; each visibility is a function of baseline and either $\nu$ or $\tau$. As noted above, we could use 
data in either frequency or delay space.  For the discussion here, we 
consider delay space and   express 
all quantities as functions of $\nu_0$. For brevity, we only retain the  time    dependence of measured visibilities.   We define: 
\begin{equation}
{\cal V} = \sum_{n=1}^N \exp(i2\pi u_0 \cos{\phi} \Delta H n) V_n
\label{eq:vis_expan_sph}
\end{equation}
The total time of over which the  visibilities are averaged  $T = N\Delta H $ should be small enough such that
the signal decorrelation is negligible (Figure~\ref{fig:HIcorr2}). For instance, we could choose $N$ such that the decorrelation is 0.9, which  corresponds roughly to 10~minutes for MWA for $\sqrt{u_0^2+v_0^2} \simeq 20$. It also follows  that if the visibilities are averaged for a period much longer than the correlation scale of the signal, there would be serious loss of the HI signal. Even though
we define  ${\cal V}$ for a single baseline $\textbf{u}_0$, it can also be obtained by averaging visibilities over all 
redundant baselines.  The correlation function that extracts the HI signal $\left |\langle V_{\tau}(\textbf{u}_0, w_{0},t)V^{*}_{\tau}(\textbf{u}_0, w_{0},t) \rangle \right |$ then is:
\begin{equation}
{\cal C}_{\rm HI} \simeq  \frac{1}{N^2} {\cal V} {\cal V}^*
\end{equation}
Notice that ${\cal C}_{\rm HI}$ is nearly the same as the expression in Eq.~(\ref{corr_def_phase}) in this case.   A longer stream of data of length, $K >>N$, can be divided into 
time slices of $N \Delta H$. The correlation function can be estimated for each slice  using this method (coherent averaging as the number of pairs is $\simeq N^2$)  and then
averaged further  over different time slices (incoherent averaging over $K/N$ slices). ${\cal C}_{\rm HI}$ is also optimal as the noise RMS is nearly the same for 
each pair of correlated  visibilities.  We note that the  HI signal is mostly contained in the real part of this resulting function, as the phase angle is small
for time scales over which the visibilities are averaged (Figure~\ref{fig:HIcorr_cov_ph}). 

A much better  method   to  utilize  the   functional form shown in Figure~\ref{fig:HIcorr2} is to use the estimator: 
\begin{equation}
{\cal C}_{\rm HI} \simeq \frac{1}{N^2} \sum_{n'} \sum_n \exp(-i2\pi u_0 \cos{\phi} \Delta H (n'-n)) V_n V_{n'}^* g^{-1}(n'-n)
\end{equation}
Here $g(n'-n)$ corresponds to the time decorrelation function shown in Figure~\ref{fig:HIcorr2}; by construction, $g(n'- n)$ is real,  $g(n - n) = 1$, and $g(n'-n)=g(n-n')$. The difference between this approach and the first method is that visibilities
are correlated first  and then averaged. This yields the same final expression
as the first method if $g(n' - n)$ is applied for a suitable time  interval such that it is close to unity.  A  distinct advantage of this method is 
that we could only retain   cross-correlations   such that
$n'\ne n$, which  allows us to avoid 
self-correlation or noise bias; the total number of cross-correlations are $\simeq N^2/2$  in this case.  This 
estimator is unbiased with respect to the detection of HI signal but does not minimize noise RMS. The following estimator is both unbiased and optimal:
\begin{equation}
{\cal C}_{\rm HI} = \frac{\sum_{n'} \sum_n \exp(-i2\pi u_0 \cos{\phi} \Delta H (n'-n)) V_n V_{n'}^* g(n'-n)} {\sum_{n'} \sum_n g^2(n'-n)}
\end{equation}
The estimator is unbiased for any choice of  $g(n' - n)$. However, for using this estimator,  small values of  $g(n'-n)$ (e.g. $g(n'-n) < 0.3$) should be avoided to prevent averaging over very noisy visibility pairs. As in the first method, the real part of this function dominates the HI signal.

The amplitude of ${\cal C}_{\rm HI}$ for  both the proposed estimators   extracts    the  visibility correlation 
function at equal time, $\langle V_{\tau}(\textbf{u}_{0}, w_{0},t)V^{*}_{\tau}(\textbf{u}_{0}, w_{0},t)\rangle$, which is real. The estimation of   HI power spectrum from this function has been extensively studied in the analysis of EoR  tracking data (e.g. \citealt{paul16}).

Our method has  similarities with  other approaches  proposed to analyze the drift scan data. In \cite{fringe-rate},   the fringe-rate filters 
have  been applied on the visibility data. We apply a similar filter to reduce 
  rapid oscillations  of the phase of the  correlation function. We note that the filter 
applied in \cite{fringe-rate} takes into all the components of earth's rotation (Eq.~(\ref{eq:angle_expand})). In our analysis, we identify the different roles
played by these components. We show how the  components responsible for the rotation and translation  of the intensity pattern cause  the decorrelation of the amplitude  of the correlation function while  the component that gives rise to  the translation  dominates the phase of the correlation function.   In $m$-mode analysis (\cite{m-mode1, m-mode2})  the intensity pattern is expanded using spherical harmonics and  the time variation of the intensity pattern is solely owing to the 
the change in the  azimuthal angle $\phi$. This time variation can then be Fourier 
transformed to extract $m$-modes of the data. The filter we apply in 
Eq.~(\ref{eq:vis_expan_sph}) corresponds to a similar  process. Eq.~(\ref{eq:vis_expan_sph}) can be viewed as a Fourier transform in which a single mode is 
extracted for a time-window of the duration  given roughly by the decorrelation time
of the amplitude  of the correlation function. Our analysis shows that 
such a procedure, directly applied on  measured visibilities, can extract the relevant information of the HI signal. 

\subsection{Impact on foregrounds}
The measured visibilities are a linear sum of 
the HI signal, foregrounds, and the noise, which are uncorrelated with each  other.  In this paper, we also compute
the time scale of the  decorrelation of a set of point source and statistically-homogeneous and isotropic diffuse foregrounds. Does our method allow us to mitigate 
foregrounds? 

First, we notice that the phase factor $\exp(-i2\pi \cos{\phi} u_0 \Delta H)$  we apply to curtail rapid oscillations of the correlation function  of  the HI signal   has the same form  for foregrounds (Eqs.~(\ref{eq:diff_corrfun}) and~(\ref{eq:forepow})).
Hence, it doesn't play a role in separating foregrounds from the HI signal. 
 
However, the decorrelation time scale of point sources is smaller  than  the HI signal. In this case, the following situation is possible: two visibilities separated in time 
are correlated  such that the HI component is fully extracted   ($g(n' - n) = 1$)
but the point source component is uncorrelated. This means that there would be   partial decorrelation  of this component of 
foregrounds  when either of the two methods discussed above are used to 
extract the HI signal.    But this argument  doesn't apply to diffuse foregrounds. 

Therefore, it is possible to partly reduce the level of foregrounds in
a drift scan  but 
the primary method of separating foregrounds from the HI signal remains transforming to delay space, as in  a  tracking observation.

\section{Summary and conclusion} \label{sec:concl}
In this paper we address the following question: over what time scales
are time-ordered visibilities coherent in a drift scan  for the EoR HI signal, set of point sources, and
 diffuse correlated foregrounds. This is an extension of our earlier work
 (\citealt{paul14}) and has similarities with other approaches in the literature
(\citealt{m-mode1,fringe-rate}). Our main theoretical tool is the complex two-point 
correlation function of visibilities measured at different times. We consider
the primary beams  of 
PAPER, MWA, HERA, and SKA1-Low for our analysis. 
Our main results can be summarized as:
\begin{itemize}
\item Figure~\ref{fig:HIcorr2} shows the amplitude of the   correlation function  of HI visibilities in time
for four interferometers. The correlation time scales  vary from a few minutes
to nearly 20~minutes for the cases considered.  We identify the three most
important factors that cause decorrelation: 
(a) traversal time  across a coherent feature, 
(b) rotation of sky intensity pattern, and 
(c) large field of view.
\item The time variation of the phase of the HI correlation function is dominated by a filter function which is    determinable in terms of measurable quantities (component of east-west baseline, latitude of the telescope, etc.).   This filter  function can be absorbed into an overall phase.  
The phase angle of the resultant function is small, which means the 
complex correlation function is dominated by its real part.  
   The  phase angle  remains small  over the coherence time scale of  the amplitude of the correlation function  (Figure~\ref{fig:HIcorr_cov_ph}). 
\item Our results are valid  in both frequency and delay space and 
are insensitive  to  the input HI power spectrum. By implication they are directly applicable to the analysis of EoR drift scan data. 
\item   The nature of foregrounds in a drift scan is different from the tracking mode owing to the time dependence of the  sky intensity pattern. We consider two components
of foregrounds for our analysis: set of  point sources and statistically 
homogeneous diffuse correlated emission. The decorrelation time scales
for these components are displayed in Figures~\ref{fig:pt_corr0203_1} and~\ref{fig:DGSE}. The point sources 
decorrelate faster than the HI signal. This provides  a novel way  to partly mitigate 
foregrounds  using only information on the sky plane. However, the  diffuse foreground decorrelation time
scale is comparable to that of the HI signal and  the contamination from this 
component cannot be removed in a drift scan on the sky plane. By implication, the delay space formalism remains  the principal method for 
isolating foregrounds from the HI signal (Figure~\ref{fig:DGSE}). 
\end{itemize}

We discuss in detail how our formalism  can be used to extract the HI signal   from the drift scan data. We argue many different approaches might be possible for the lossless retrieval of the HI signal  while optimizing the noise. In the future, we 
hope to apply our formalism  to  publicly-available  drift scan data.

\appendix
\section{Coordinate Transformation} \label{app:coord}
Here we discuss sky coordinate system ($l,m,n$) in terms of $(\delta, \phi, H)$ with $\delta, \phi, H$ representing the declination, the  terrestrial latitude of the telescope, and the  hour angle, respectively. From Eq.~(A4.7) of \cite{ChristiansenHoegbom}:
\begin{align}
	\begin{split}
	l &= \cos{\delta} \sin{H} 										\\	
	m &= \cos{\delta} \cos{H} \sin{\phi} - \sin{\delta} \cos{\phi} 	\\	
	n &= \cos{\delta} \cos{H} \cos{\phi} + \sin{\delta} \sin{\phi} 	\\	
	\end{split}
\end{align}
In a drift scan, the primary beam remains  unchanged  with respect to  a fixed phase center chosen to be 
$l=m=0$.  The  coordinates  of intensity pattern  ($l,m,n$) change with time, in the first 
order in $\Delta H$, as:
\begin{align}\label{eq:dldmdn}
	\begin{split}
		\Delta l &= \left(m \sin{\phi} + n \cos{\phi} \right) \Delta H \\
		\Delta m &= -l \sin{\phi} \Delta H \\
		\Delta n &= -l \cos{\phi} \Delta H
	\end{split}
\end{align}
The change in hour angle, $\Delta H$,  can be expressed in terms of radians as: 
  \begin{align} \label{eq:dhdt}
   \Delta H [\text{in rad}] = \frac{\pi}{12} \frac{\Delta t [\text{in min}]}{60}
  \end{align}

We use Eq.~(\ref{eq:dldmdn}) to express the time-dependent part of Eq.~(\ref{eq:viscorapprox2}) explicitly in terms of change in hour angle $\Delta H$. Eq.~(\ref{eq:dhdt}) can be used to express $\Delta H$ in terms of drift time $\Delta t$ for a zenith scan.
\begin{align}\label{eq:angle_expand}
 -\frac{r_{0}}{2\pi} \textbf{k}_{\perp}\cdot \Delta {\boldsymbol{\vartheta}(\Delta t)}  &= -\frac{r_{0}}{2\pi} \left( k_{\perp 1} \Delta l + k_{\perp 2} \Delta m \right) \nonumber \\
  &= -\frac{r_{0}}{2\pi} \left( k_{\perp 1}\left(m \sin{\phi} + n \cos{\phi} \right) \Delta H - k_{\perp 2}l \sin{\phi} \Delta H \right)  \nonumber \\
  &\simeq -\frac{r_{0}}{2\pi}  \left( k_{\perp 1} \cos{\phi} \Delta H + \left( -l k_{\perp 2} + m k_{\perp 1} \right) \sin{\phi} \Delta H \right) + \frac{1}{2} \left( l^{2} + m^{2} \right) \frac{r_{0}}{2\pi}  k_{\perp 1} \cos{\phi} \Delta H  
 \end{align}
We use  the flat-sky approximation $n \simeq 1 -\frac{1}{2} \left(l^{2} + m^{2}\right)$ in writing Eq.~(\ref{eq:angle_expand}).

\section{Further simplification of visibility correlation function}

In this appendix we discuss how the visibility correlation function can be 
further simplified for large primary beams and long baselines. This allows us to discern several
generic properties of the correlation function. 
We start with the HI visibility correlation function in frequency space (Eq.~(\ref{eq:viscorapprox2})):
\begin{align}
	\Big\langle V_{\nu}(\textbf{u}_{\nu}, w_{\nu},t) V^{*}_{\nu'}(\textbf{u}'_{\nu'}, w'_{\nu'},t') \Big\rangle = 
	\bar{I}_{\nu} \bar{I}_{\nu'} {\int} \frac{d^{3}k}{(2\pi)^{3}} P_{HI}(k) e^{i k_{\parallel} |\dot{r}_{0}| \Delta \nu} \nonumber \\
	& \mkern-400mu	e^{i r_{\nu} k_{\perp 1} \cos{\phi} \Delta H}
	Q_\nu(\textbf{k}_{\perp}, \textbf{u}_{\nu}, w_{\nu}, \Delta H = 0) Q^*_{\nu'}(\textbf{k}_{\perp}, \textbf{u}'_{\nu'}, w'_{\nu'}, \Delta H)	\nonumber
\end{align}

The  Fourier beam can be expressed as (Eq.~(\ref{eq:viscorapprox3})):
\begin{align}
	 Q_\nu(\textbf{k}_{\perp}, \textbf{u}_{\nu}, w_{\nu}, \Delta H) = {\int} d^{2}\theta A_{\nu}(\boldsymbol{\theta}) 
	\exp\left[ -2 \pi i \left( \boldsymbol{x}_u \cdot {\boldsymbol{\theta}} -\frac{1}{2} y \theta^{2}\right) \right] \label{eq:fourier_beam}
\end{align}
	with
\begin{align*}
	x_u &= u_{\nu} - \frac{r_{\nu}}{2\pi} \left(k_{\perp 1} + k_{\perp 2} \sin{\phi} \Delta H \right) \\
	x_v &= v_{\nu} - \frac{r_{\nu}}{2\pi} \left(k_{\perp 2} - k_{\perp 1} \sin{\phi} \Delta H \right) \\
	y   &= w_{\nu} + \frac{r_{\nu}}{2\pi} k_{\perp 1} \cos{\phi} \Delta H
\end{align*}

We consider  a Gaussian beam: $A(l,m) =  e^{-(l^2 + m^2)/\Omega_{g}} $ to compute the 
Fourier beam: 
\begin{align*}
Q_\nu(\textbf{k}_{\perp}, \textbf{u}_{\nu}, w_{\nu}, \Delta H) = Q(x_u, x_v, y) = \frac{\pi \Omega_{g}} {1 - i \pi y \Omega_{g} } \exp \left[ - \frac{\pi^2 \Omega_{g} (x_u^2 + x_v^2) } {1 - i \pi y \Omega_{g}} \right]
\end{align*}
For  $\Omega'_{g} \equiv \Omega_{g} / (1 - i \pi y \Omega_{g} ) $
\begin{align*}
Q_\nu(\textbf{k}_{\perp}, \textbf{u}_{\nu}, w_{\nu}, \Delta H) = Q(x_u, x_v, y) = \pi \Omega'_{g} \exp \left[ - \pi^2 \Omega'_{g} (x_u^2 + x_v^2) \right]
\end{align*}
If $\Omega_g$ is large, e.g. PAPER or MWA beams, we can use $\delta$-function approximation for solving $Q_\nu(\textbf{k}_{\perp}, \textbf{u}_{\nu}, w_{\nu}, \Delta H = 0)$, which gives us: 
\begin{align*}
Q_\nu(\textbf{k}_{\perp}, \textbf{u}_{\nu}, w_{\nu}, \Delta H = 0) &= 
\delta \left(u_{\nu} - \frac{r_{\nu}}{2\pi} k_{\perp 1} \right) \delta \left(v_{\nu} - \frac{r_{\nu}}{2\pi} k_{\perp 2} \right) \\
Q_\nu(\textbf{k}_{\perp}, \textbf{u}_{\nu}, w_{\nu}, \Delta H = 0) &= 
 \left( \frac{2 \pi}{r_\nu} \right)^2 \delta^2 \left( \textbf{k}_{\perp} - \frac{2\pi}{r_{\nu}} \textbf{u}_{\nu} \right)
 \end{align*}

This allows us to express HI visibility correlation function in frequency space as:
\begin{align}
	\Big\langle V_{\nu}(\textbf{u}_{\nu}, w_{\nu},t) V^{*}_{\nu'}(\textbf{u}'_{\nu'}, w'_{\nu'},t') \Big\rangle = 
	\frac{ \bar{I}_{\nu} \bar{I}_{\nu'} } {r^2_\nu} e^{ 2 \pi i u_{\nu} \cos{\phi} \Delta H}
	Q^*_{\nu'}(\textbf{k}_{\perp}, \textbf{u}'_{\nu'}, w'_{\nu'}, \Delta H) \nonumber \\
	& \mkern-250mu
	{\int} \frac{d k_{\parallel}} {2\pi} P_{HI}(k) e^{i k_{\parallel} |\dot{r}_{0}| \Delta \nu} \label{eq:viscorr_delta}
\end{align}
In the previous equation we have used, $\textbf{k}_{\perp} = {2\pi}  \textbf{u}_{\nu} / {r_{\nu}}$.
Eq.~(\ref{eq:viscorr_delta}) gives an excellent approximation for MWA and PAPER, and for HERA and SKA1-Low for long baselines  in frequency space. This can be readily be computed 
at any frequency and  explains  the features seen in Figure~\ref{fig:HIcorr2}. 

We  can extend our analysis  to HI visibility correlation function in delay space (Eq.~(\ref{eq:visdelaysp})):
\begin{align*}
	\Big\langle V_{\tau}(\textbf{u}_{0}, w_{0},t)V^{*}_{\tau}(\textbf{u}'_{0}, w'_{0},t') \Big\rangle 
	 = \iint _{\nu_0 - B/2} ^{\nu_0 + B/2} d\nu d\nu' \Big\langle V_{\nu}(\textbf{u}_{\nu}, w_{\nu},t) V^{*}_{\nu'}(\textbf{u}'_{\nu'}, w'_{\nu'},t') \Big\rangle 
	 e^{-2 \pi i \tau \Delta \nu }
\end{align*}
Here $B$ is the observational bandwidth. We make the same approximations discussed in section~\ref{sec:delay_space}, which gives us:
\begin{align*}
	\Big\langle V_{\tau}(\textbf{u}_{0}, w_{0},t)V^{*}_{\tau}(\textbf{u}'_{0}, w'_{0},t') \Big\rangle 
	 = 	\frac{ \bar{I}^2_{0} } {r^2_0} e^{ 2 \pi i u_{0} \cos{\phi} \Delta H}
	Q^*_{\nu_0}(\textbf{k}_{\perp}, \textbf{u}'_{0}, w'_{0}, \Delta H) \\
	& \mkern-250mu {\int} \frac{d k_{\parallel}} {2\pi} P_{HI}(k) 
	\iint _{\nu_0 - B/2} ^{\nu_0 + B/2} d\nu d\nu'	e^{i \Delta \nu ( k_{\parallel} |\dot{r}_{0}| - 2 \pi  \tau ) } \\
	\Big\langle V_{\tau}(\textbf{u}_{0}, w_{0},t)V^{*}_{\tau}(\textbf{u}'_{0}, w'_{0},t') \Big\rangle 
	 \simeq 	\frac{ \bar{I}^2_{0} } {r^2_0} e^{ 2 \pi i u_{0} \cos{\phi} \Delta H}
	Q^*_{\nu_0}(\textbf{k}_{\perp}, \textbf{u}'_{0}, w'_{0}, \Delta H) \\
	& \mkern-250mu {\int} \frac{d k_{\parallel}} {2\pi} P_{HI}(k) 
	\frac{2\pi B}{|\dot{r}_{0}|} \delta \left( k_{\parallel} - \frac{2\pi \tau}{|\dot{r}_{0}|} \right)
 \end{align*}
 In deriving this equation,   we use the following result
from section~\ref{sec:delay_space}: 
\begin{align*}
\iint _{\nu_0 - B/2} ^{\nu_0 + B/2} d\nu d\nu'	e^{i \Delta \nu ( k_{\parallel} |\dot{r}_{0}| - 2 \pi  \tau ) } = 
B^{2} \sinc^{2}\left[\pi B \left( \tau - \frac{|\dot{r}_{0}|}{2\pi} k_{\parallel} \right) \right ] \simeq 
\frac{2\pi B}{|\dot{r}_{0}|} \delta \left( k_{\parallel} - \frac{2\pi \tau} {|\dot{r}_{0}|} \right) \label{eq:delfnapprox}
\end{align*}
The HI signal is strongly correlated when 
$| \textbf{u}_{0} - \textbf{u}'_{0} | \lesssim 2/\Omega_g^{1/2}$, which allows us to use  $\textbf{u}'_{0} \approx \textbf{u}_{0} $. This gives us:
\begin{equation}
\Big\langle V_{\tau}(\textbf{u}_{0}, w_{0},t)V^{*}_{\tau}(\textbf{u}'_{0}, w'_{0},t') \Big\rangle \simeq 	\frac{ \bar{I}^2_{0} B } {r^2_0 |\dot{r}_{0}| } e^{ 2 \pi i u_{0} \cos{\phi} \Delta H}
	Q^*_{\nu_0}(\textbf{k}_{\perp}, \textbf{u}_{0}, w'_{0}, \Delta H) P_{HI}(k)
\label{eq:viscorr_delta1}
\end{equation}
where $ k = \sqrt{ \left( {2\pi} \tau / {|\dot{r}_{0}|}\right)^2 + 
				   \left( {2\pi} u_{0} /{r_{0}} \right)^2 + 
				   \left( {2\pi} v_{0} /{r_{0}}  \right)^2 	}$. 
Though Eq.~(\ref{eq:viscorr_delta1}) was derived  using a Gaussian beam, it is in excellent agreement  with the numerical results for MWA and PAPER and 
for HERA and SKA1-Low for longer baselines ($|\textbf{u}| \ga 150$)  shown in Figure~\ref{fig:HIcorr2}. Eq.~(\ref{eq:viscorr_delta1}) also shows that the decorrelation time  is expected to be nearly independent of the  delay parameter   $\tau$.

We next give  explicit forms of the amplitude and the phase of the Fourier beam. We have:
\begin{align*}
Q_\nu(\textbf{k}_{\perp}, \textbf{u}_{\nu}, w_{\nu}, \Delta H) = Q(x_u, x_v, y) = \frac{\pi \Omega_{g}} {1 - i \pi y \Omega_{g} } \exp \left[ - \frac{\pi^2 \Omega_{g} (x_u^2 + x_v^2) } {1 - i \pi y \Omega_{g}} \right]
\end{align*}
where $x_u^2 + x_v^2 = | \textbf{u}_{\nu} |^2 \sin^2{\phi} \Delta H ^2 $ and 
	  $y = w_{\nu} + u_{\nu} \cos{\phi} \Delta H $. Then,
\begin{align*}
Q_\nu(\textbf{k}_{\perp}, \textbf{u}_{\nu}, w_{\nu}, \Delta H) &= \pi z_1 z_2 = \pi a_1 e^{i \psi_1} a_2 e^{i \psi_2} 
		  = \pi a_1 a_2 e^{i (\psi_1 + \psi_2) } \\
 {\rm Amp} \left[ Q_\nu(\textbf{k}_{\perp}, \textbf{u}_{\nu}, w_{\nu}, \Delta H) \right] &=  \pi a_1 a_2	\\
 {\rm Arg} \left[ Q_\nu(\textbf{k}_{\perp}, \textbf{u}_{\nu}, w_{\nu}, \Delta H) \right] &=  \psi_1 + \psi_2 \\
z_1 &= a_1 e^{i \psi_1} = \frac{ \Omega_{g}} {1 - i \pi y \Omega_{g} }   \\
z_2 &= a_2 e^{i \psi_2} = \exp \left[ - \frac{\pi^2 \Omega_{g} (x_u^2 + x_v^2) } {1 - i \pi y \Omega_{g}} \right]
\end{align*}
On solving $a_1, \psi_1, a_2, \psi_2$ in terms of known quantities, we find;
\begin{align*}
	a_1 &= \frac{\Omega_g}{ \sqrt{ 1 + \pi^2 \Omega_g^2 y^2 } } \\
 \psi_1 &= \arctan{ (\pi \Omega_{g} y) } \\
	a_2 &= \exp \left[ - \pi^2 (x_u^2 + x_v^2) a_1 \cos{\psi_1} \right] 
				= \exp \left[ - \pi^2 (x_u^2 + x_v^2) \frac{\Omega_g}{ 1 + \pi^2 \Omega_g^2 y^2 } \right]\\
 \psi_2 &= -\pi^2 (x_u^2 + x_v^2) a_1 \sin{\psi_1} 
				= -\pi^2 (x_u^2 + x_v^2) \frac{\Omega_g}{ 1 + \pi^2 \Omega_g^2 y^2 } (\pi \Omega_{g} y)
\end{align*}
Hence,
\begin{align}
 {\rm Amp} \left[ Q_\nu(\textbf{k}_{\perp}, \textbf{u}_{\nu}, w_{\nu}, \Delta H) \right] &=  \pi a_1 a_2	
	  =	\frac{\pi \Omega_g}{ \sqrt{ 1 + \pi^2 \Omega_g^2 y^2 } } 
			\exp \left[ - \pi^2 (x_u^2 + x_v^2) \frac{\Omega_g}{ 1 + \pi^2 \Omega_g^2 y^2 } \right] \nonumber	\\
 {\rm Arg} \left[ Q_\nu(\textbf{k}_{\perp}, \textbf{u}_{\nu}, w_{\nu}, \Delta H) \right] &=  \psi_1 + \psi_2 
	  = \arctan{ (\pi \Omega_{g} y) } -\pi^2 (x_u^2 + x_v^2) \frac{\Omega_g}{ 1 + \pi^2 \Omega_g^2 y^2 } (\pi \Omega_{g} y)  \label{eq:phase_gua}
\end{align}
The total phase acquired by the HI visibility correlation function is  $ 2 \pi u_{0} \cos{\phi} \Delta H + \psi_1 + \psi_2 $.


\end{document}